\documentclass[10pt, journal, compsoc]{IEEEtran}

\usepackage{algorithm}
\usepackage{algorithmic}
\usepackage{amsfonts}
\usepackage{amsmath}
\usepackage{amssymb}
\usepackage{booktabs}
\usepackage{cite}
\usepackage{float}
\usepackage{footnote}
\usepackage{graphicx}
\usepackage{mathrsfs}
\usepackage{multirow}
\usepackage{stfloats}
\usepackage{stfloats}
\usepackage{subfigure}
\usepackage{color}

\usepackage{url}
\usepackage[hyphenbreaks]{breakurl}

\newtheorem{theorem}{Theorem}
\newtheorem{lemma}{Lemma}

\begin{document}

\title{Online User-AP Association with Predictive Scheduling in Wireless Caching Networks}

\author{
Xi~Huang,~\IEEEmembership{Student Member,~IEEE}, 
Shuang Zhao, 
Xin~Gao,~\IEEEmembership{Student Member,~IEEE},\\
Ziyu~Shao$^*$,~\IEEEmembership{Senior Member,~IEEE}, 
Hua~Qian,~\IEEEmembership{Senior Member,~IEEE},
Yang~Yang,~\IEEEmembership{Fellow,~IEEE}
\thanks{
{\indent 
		$^*$ The corresponding author of this work is Ziyu Shao. 
		
		X. Huang, X. Gao, Z. Shao, and Y. Yang are with the School of Information Science and Technology, ShanghaiTech University, China. (E-mail:\{huangxi, gaoxin, shaozy, yangyang\}@shanghaitech.edu.cn)

		S. Zhao was with Shanghai Institute of Microsystem and Information Technology, Chinese Academy of Sciences and University of Chinese Academy of Sciences, China. (E-mail: nicholezhao@tencent.com)}
				
		H. Qian is with the University of Chinese Academy of Sciences and Shanghai Advanced Research Institute, Chinese Academy of Sciences, China. (E-mail: qianh@sari.ac.cn)
}%
}

\IEEEtitleabstractindextext{
\begin{abstract}
For wireless caching networks, the scheme design for content delivery is non-trivial in the face of the following tradeoff. On one hand, to optimize overall throughput, users can associate their nearby APs with great channel capacities; however, this may lead to unstable queue backlogs on APs and prolong request delays. On the other hand, to ensure queue stability, some users may have to associate APs with inferior channel states, which would incur throughput loss. Moreover, for such systems, how to conduct predictive scheduling to reduce delays and the fundamental limits of its benefits remain unexplored. In this paper, we formulate the problem of online user-AP association and resource allocation for content delivery with predictive scheduling under a fixed content placement as a stochastic network optimization problem. By exploiting its unique structure, we transform the problem into a series of modular maximization sub-problems with matroid constraints. Then we devise \textit{PUARA}, a Predictive User-AP Association and Resource Allocation scheme which achieves a provably near-optimal throughput with queue stability. Our theoretical analysis and simulation results show that PUARA can not only perform a tunable control between throughput maximization and queue stability, but also incur a notable delay reduction with predicted information.
\end{abstract}

}

\maketitle
\IEEEdisplaynontitleabstractindextext
\IEEEpeerreviewmaketitle

\IEEEraisesectionheading{\section{Introduction}\label{sec:introduction}}
\IEEEPARstart{R}{ecent} years have witnessed an explosive growth of global mobile data traffic \cite{CiscoIndex}. 
To accommodate the ever-increasing traffic, a number of techniques such as network densification have been proposed to be applied in 5G networks to increase network capacity
\cite{andrews2014will5G,shi2014group,wang2016mobility}. 
However, such techniques may also impose heavy burdens on backhaul links for wireless access at the network edge, thereby leading to a significant degradation in quality-of-service (QoS). 
To mitigate such issues, wireless caching has come as a promising solution to promote QoS and ease the burden of backhaul links \cite{liu2016caching}.
The key idea is to pre-fetch or cache frequently requested contents (\textit{e.g.}, news feed and video streaming) onto edge devices, \textit{e.g.}, small-cell wireless access points (APs)
\cite{shanmugam2013femtocaching,liu2016contentCachingICC,Peng2015Globalcom,Peng2016ICCCacheSizeAllocation} 
and user terminals
\cite{Golrezaei2013femtocachingMagzine}\cite{ji2016wirelessD2D}. 

Basically, there are two critical phases in wireless caching networks\cite{maddah2014fundamental}. 
The first phase, \textit{a.k.a.} \emph{content placement phase}, focuses on how to distribute contents efficiently on APs with limited caching resources. 
The second phase, \textit{a.k.a.} \emph{content delivery phase}, considers how to determine user-AP association in a dynamic fashion, within which requested contents are delivered from APs to users.  
Considering the costs of cache update and content migration, content placement can only be conducted infrequently. 
Therefore, content placement often proceeds at a larger time scale than user-AP association and can be viewed as a static operation.

Regarding content delivery, so far, it is still an open problem on how to design effective user-AP association schemes so as to minimize delivery latencies and maximize various network utilities such as throughput.
The key challenge comes from that the statistics of user request traffic and wireless dynamics are usually unattainable in practice.  
All such uncertainties make it difficult to conduct effective content delivery in an online fashion. Moreover, it is often desirable to yield a design with performance guarantee, 
so that system designers are well aware of how far the system proceeds away from the optimal performance and how to further improve the system under various design tradeoffs. 

Motivated by the rapid development of machine learning in recent years, there has been a growing trend among various practical systems in exploiting the short-term predictability of network traffic or user behaviors to conduct proactive pre-service to promote system performance. 
For example, Netflix prefetches videos to users based on user behavior prediction \cite{NetflixPred}.
Inspired by the wide adoption of prediction-based approaches \cite{nanda2016predicting, zhang2017proactive, petrov2018mathematical, huang2019streaming, huang2019sdn, huang2019nfv, gao2019fog}, 
we pose an interesting question: 
in wireless caching networks, if the users' requests for content delivery can be effectively predicted and pre-served, even within a short time window ahead of its actual arrival, 
then what are the \textit{fundamental} benefits of such predictive scheduling and the impacts of prediction errors? 
To date, there is still a lack of systematic investigations on the answer to such a question. 
Such investigations can serve as the basis to understand the endeavor worthy to be put on incorporating predictive scheduling into wireless caching networks and the costs that we can afford in the worst case.  

In this paper, we study the problem of joint user-AP association and resource allocation in resource-limited wireless caching networks with fixed content placements. 
By proposing a predictive and online scheme to solve such a problem, 
we conduct a systematic study to investigate the fundamental benefits of predictive scheduling in wireless caching networks with both theoretical analysis and experimental verification.
We summarize our key results and main contributions as follows.
\begin{itemize}
  \item[$\diamond$] \textbf{System Modeling and Problem Formulation:} 
  We develop a novel system model by taking a careful choice in the granularities of system state characterization and the decision-making procedure.
  By leveraging the idea of predictive window techniques in \cite{Huang2014When}, our model captures the dynamics of predictive scheduling at the granularity of file units while depicting the decision-making procedure on a per-time-slot basis to mitigate the online control overheads. 
  Then we formulate the problem of user-AP association and resource allocation as a stochastic network optimization problem, with the aim to maximize the long-term time-average overall network throughput and achieve the stability of all queues in the system.
  \item[$\diamond$] \textbf{Algorithm Design:} 
  To solve the formulated problem in the face of time-varying wireless environment dynamics and unknown user traffic statistics, we take a non-trivial transformation to convert the long-term stochastic optimization problem into a series of sub-problems over time slots. 
  By exploiting the unique structure of such sub-problems, we  reformulate each of them as a modular maximization problem over two intersected matroid constraints. 
  Then we propose \textit{PUARA}, an effective \textit{Predictive User-AP Association and Resource Allocation} scheme, which achieves a tunable control between throughput maximization and queue stabilization, while taking advantage of predicted information to reduce request sdelays.
  \item[$\diamond$] \textbf{Performance Analysis:} 
  	We conduct theoretical analysis to evaluate the performance of PUARA. Our results show that PUARA can achieve a near-optimal time-average network throughput with an approximation ratio of $1/2$ while guaranteeing the queue stability in the system. 
  	Moreover, by leveraging predicted information, PUARA can achieve even better performance with a notable delay reduction proportional to the prediction window sizes.
  \item[$\diamond$] \textbf{Experimental Verification:}
  	We conduct extensive simulations to evaluate the performance of PUARA and explore the fundamental limits of the benefit of predictive scheduling in \textit{delay reduction} for wireless caching networks. Besides, we also evaluate the performance of PUARA under two kinds of mis-prediction, \textit{i.e.}, the mis-prediction of the \textit{type} and the \textit{size} of requested files. Our results demonstrate the robustness of PUARA against mis-predictions.
  \item[$\diamond$] \textbf{Predictive Scheduling:}
  To our best knowledge, this paper is the first \textit{systematic} study to explore the \textit{fundamental} limits of the benefit of predictive scheduling on delay reduction in wireless caching networks with both performance analysis and simulations. Our results provide novel insights to improve the design of wireless caching networks.
\end{itemize}


The rest of this paper is organized as follows. 
Section \ref{sec: related work} discusses the related work. 
In Section \ref{Sys}, we present our system model and problem formulation. 
Next, Section \ref{PUARA} shows the design of PUARA, followed by the corresponding performance analysis.
Then Section \ref{Simulation} presents our simulation results and corresponding discussion, while Section \ref{Conclusions} concludes this paper.

\section{Related Work} \label{sec: related work}

\textbf{{Content placement:}}
For wireless caching networks, most existing works focus on the content placement problem.
For example, Shanmugam \emph{et al.} \cite{shanmugam2013femtocaching} developed two effective cache placement strategies to minimize the expected file download time for streaming services upon a fixed network topology (user-AP connectivity).
Peng \emph{et al.} \cite{Peng2015Globalcom} further considered the delay over backhaul links induced by file request transmission due to cache miss into account. 
They then proposed a centralized scheme which leverages base station cooperation to optimize the cache placement.
Later, Song \emph{et al.} \cite{song2017optimal} further took the impacts of wireless fading into account. By revealing the tradeoff between file diversity gain and channel diversity gain, they developed a greedy content placement scheme to minimize the average bit error rate (BER) in wireless networks under Rayleigh flat fading channel model.
Unlike such a line of works which generally require centralized coordination with full knowledge of system dynamics, 
another line of works have considered optimizing the placement in a \textit{distributed} manner.   
For instance, Liu \textit{et al.} \cite{liu2016contentCachingICC} considered the caching placement problem in dense network settings. 
With the aim to minimize the expected delay for serving user requests, they formulated the problem as a constrained integer programming problem and proposed a belief-propagation-based scheme to decide the caching placement in a distributed manner.
Zheng \textit{et al.} \cite{zheng2018stackelberg} studied the content placement problem in large-scale mobile edge networks from a game-theoretic perspective. 
By formulating the problem as a Stackelberg game, they proposed a framework for scalable and convergent incentive mechanism design for edge caching. Their solution allows the content placement to be decided in a distributed fashion through limited interactions between edge nodes and users.

In parallel, some other recent works further consider conducting content placement based on learned statistics of user mobility or user demands such as the distribution of content popularity to optimize the quality of services, \textit{a.k.a.} proactive caching. 
For example, Vasilakos \textit{et al.} \cite{vasilakos2016addressing} considered leveraging predicted user mobility to minimize the expected costs incurred by the delivery of personalized and dynamic content.
Meanwhile, M{\"u}ller \textit{et al.} \cite{muller2016context} developed a proactive caching scheme which learns context-specific content popularity and proactively updates the cached content in an online fashion. 
Likewise, Doan \textit{et al.} \cite{doan2018content} leveraged feature clustering techniques to foresee the content popularity and optimized the content placement based on such predictions.
Later, Chen \textit{et al.} \cite{chen2017echo} developed another effective scheme which jointly learns the distribution of user request demand and user mobility pattern to optimize the cache placement with the best QoS. 

Our work is orthogonal to the above works since our focus is on optimizing the online content delivery between users and APs upon a fixed content placement. 
In fact, our solution can also be integrated effectively with them to optimize the performance of wireless caching networks.

\textbf{Content delivery:}
To date, there has also been a number of works proposed with respect to content delivery in wireless caching networks\cite{liu2016caching}. 
Particularly, we focus on the works which exploit predictive scheduling to optimize content delivery. 

Most of such works leveraged predictive scheduling (\textit{a.k.a.} \textit{anticipatory scheduling} \cite{draxler2013anticipatory,sadr2013anticipatory,bui2015anticipatory,draxler2015smarterphones,Bethanabhotla2015videoStreaming,tsilimantos2016anticipatory}) to improve the quality of video streaming services in wireless caching networks.
For example, Sadr \textit{et al.} \cite{sadr2013anticipatory} studied the single-user buffer control problem for video streaming services. 
By assuming the future channel states can be perfectly predicted in a finite lookahead time window, they proposed a time-slot-based scheduling scheme which pre-allocates wireless channel resources and buffers to minimize the fraction of required bandwidth to meet the user's QoS. 
Meanwhile, Dr{\"a}xler \textit{et al.} \cite{draxler2013anticipatory} considered the joint buffer control and quality selection problem under a multi-user setting. By assuming that data rates are perfectly predicted, they developed two heuristic schemes to plan the quality and download time of video segments to eliminate playback interruptions.
Later, Bui \textit{et al.} \cite{bui2015anticipatory} explored the  benefits of perfect system state prediction for media streaming in mobile networks. 
Based on their mixed-integer linear programming problem formulation, they developed a heuristic scheme to perform predictive admission control and resource allocation to maximize availability of streaming services to users.
In parallel, Dr{\"a}xler \textit{et al.} \cite{draxler2015smarterphones} took a further step by considering more general settings with imperfect data rate prediction. They proposed another heuristic scheme to optimize both the quality and download time of video services. 
These works, with their justified effectiveness in exploiting predicted information to improve resource (\textit{e.g.}, buffers and bandwidth) utilization, are explicitly designed for streaming services and generally provide no theoretical performance guarantee. 

Different from existing works, our work focuses on the content delivery for \textit{general} multi-user multi-AP wireless caching networks. Moreover, our proposed schemes can jointly optimize the throughput and queue stability for such systems in a tunable fashion with theoretical performance guarantee. 
Besides, we also investigate the \textit{fundamental} limits of the benefit of predictive scheduling in \textit{delay reduction} based on the prediction of user request traffic, as well as the impacts of the mis-predictions of requested files' type and size. 
To our best knowledge, our work is the first systematic study with extensive theoretical and experimental results to characterize such limits.

 \begin{figure}[!t]
    \centering{
    \includegraphics[width=0.85\columnwidth]{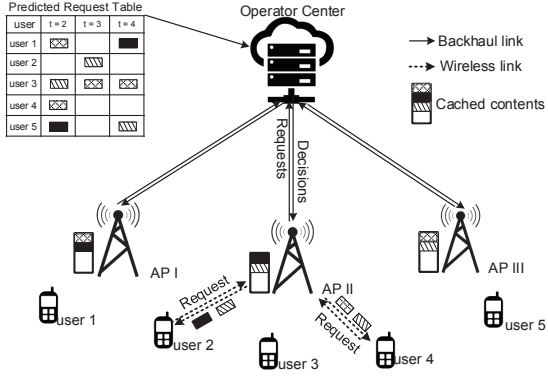}}
    \caption{An example of cache-enabled content-centric wireless networks.}
    \label{fig:SystemModel}
 \end{figure}
 \setlength{\floatsep}{0pt}

\section{System Model and Problem Formulation}\label{Sys}
In this section, we demonstrate our system model and problem formulation in detail.
For ease of understanding, we show an instance of our system model in Figure \ref{fig:SystemModel} and 
summarize the key notations in TABLE \ref{tab:Notation}.

\subsection{Basic System Settings}
We consider a cache-enabled content-centric wireless network (CCWN) which serves a number of users (denoted by set $\mathcal{U}$ with $|\mathcal{U}|=U$) through a set of APs (denoted by set $\mathcal{H}$ with $|\mathcal{H}|=H$).
The system proceeds over time slots that are indexed by $t \in \{0,1,2,\ldots\}$ and each of them has a fixed length of $\tau$.
Particularly, there is an operation center that connects to all APs via backhaul links. 
The operation center maintains all files that could be requested by users, which are grouped into $F$ types by their corresponding services. 
We denote the set of all file types by $\mathcal{F}$.   
Due to resource limitations, each AP only caches $N$ types of files ($N < F$). The placement of cached files across APs is assumed to be fixed and pre-determined by existing schemes such as\cite{muller2016context,song2017optimal,chen2017echo,doan2018content,Ji2013optimal,ji2015throughput,Ji2016fundamental,Psaras2014Information-centric,quer2018proactive}.\footnote{Our model can serve as a basis to consider more general scenarios in which the number of cached file types is different across APs.}
In the meanwhile, users are distributed around APs and send requests for retrieving files from such APs via wireless links. 
Particularly, for each type $f$ of files, each user $u$ maintains a local queue to buffer their corresponding unserved requests.
We use $Q_{uf}(t)$ to denote the total size of unserved files which are of type $f$ and requested by user $u$ within time slot $t$. 
For ease of notation, we define $\boldsymbol{Q}(t) \triangleq (Q_{uf}(t),u \in \mathcal {U}, f \in \mathcal {F})$ and assume that all queue backlogs are initially empty, \textit{i.e.},
	$Q_{uf}(0)= 0$, for each user $u \in \mathcal {U}$ and each file type $f \in \mathcal {F}$.

During each time slot $t$, the system proceeds as follows. 
At the beginning of the time slot, 
the operation center collects instant system dynamics such as different users' queue backlog sizes and wireless channel dynamics, then decides the user-AP association and service rate allocation.
Based on such decisions, each AP associates with a subset of users and decides the service rate allocation among the users.
Then each user retrieves its requested files based on such allocations.\footnote{
		Note that file download is viable only when the requested type of files have been cached on their associated AP.} 
At the end of the time slot, each user appends its new file requests to its local queues by their types, then updates its local queue sizes accordingly.

\begin{table}[!t]
    \caption{Key notations in the system model}
    \begin{center}
        \begin{tabular}{ll}
          \hline\hline
          \textbf{Notation} & \textbf{Description} \\
          \hline\hline
          $\mathcal {U}$ & Index set of the users \\
          \hline
          $\mathcal {H}$ & Index set of the APs \\
          \hline
          $\mathcal {F}$ & Index set of the files \\
          \hline
          $\mathcal {E}$ & user-AP potential link set \\
          \hline
		  $\tilde{\mathcal {E}}$ & AP-file association set \\
          \hline
          \multirow{2}*{$M$} & Maximum number of associated users for each AP \\
              & in one time slot \\
          \hline
          \multirow{2}*{$X_{uh}(t)$} & Association indicator for user $u$ and AP $h$ \\
          & in time slot $t$ \\
          \hline
          $Y_{hf}$ & Association indicator AP $h$ and file type $f$ \\
          \hline
          $\boldsymbol {X}(t)$ & User-AP association matrix \\
          \hline
          $\boldsymbol {Y}$ & AP-file association matrix \\
          \hline
          $A_u(t)$ & Requested files amount by user $u$ in time slot $t$ \\
          \hline
          $\boldsymbol {A}(t)$ & Request arrival vector \\
          \hline
          \multirow{2}*{$I_{uf}(t)$} & Indicator for whether user $u$ requests file type $f$ \\
          		& in time slot $t$ \\
          \hline
          $\boldsymbol {I}(t)$ & Indicator matrix for user requests \\
          \hline
          \multirow{2}*{$C_{uh}(t)$} & Maximum achievable rate over link $(u,h)$ \\
                      &  in time slot $t$ \\
          \hline
          \multirow{2}*{$\nu_{uh}(t)$} & Allocated bandwidth proportion for user $u$ from \\
          & AP $h$ in time slot $t$ \\
          \hline
          \multirow{2}*{$\mu_{uh}(t)$} & Allocated service rate for user $u$ from \\
          & AP $h$ in time slot $t$ \\
          \hline
          $\mu_{uf}(t)$ & Service rate for downloading file $f$ in time slot $t$ \\
          \hline
          \multirow{2}*{$\tilde{\mu}_{uf}^{d}(t)$} & Allocated service rate for arriving request for file \\
              & type $f$ in time slot $(t+d)$ \\
          \hline
          \multirow{2}*{$\tilde{Q}_{uf}^{d}(t)$} & Queue backlog of untreated requests on user $u$ with \\
          & respect to file type $f$ in $d$ slots ahead of time $t$ \\
          \hline
        \end{tabular}
    \end{center}
    \label{tab:Notation}
  \end{table}

\subsection{User-AP Association and Cache Placement}
We model the user-AP association in the caching network as a bipartite graph $\mathcal {G} = (\mathcal {U},\mathcal {H},\mathcal {E})$. 
Each edge $(u, h)$ in the edge set $\mathcal{E}$ of the graph indicates that there exists a potential transmission link between AP $h \in \mathcal {H}$ and user $u \in \mathcal {U}$.
 
For each time slot $t$, we define $\boldsymbol{X}(t) \in \{0, 1\}^{U \times H}$ as the matrix that describes the actual user-AP association with respect to $\mathcal {G}$. 
For each user-AP pair $(u,h) \in \mathcal{E}$, $X_{uh}(t) = 1$ implies that user $u$ and AP $h$ are associated within time slot $t$ and zero otherwise.
If there is no potential link between user $u$ and AP $h$ (\textit{i.e.}, $(u,h) \notin \mathcal {E}$), we always have $X_{uh}(t) = 0$. 

For each AP $h\in \mathcal {H}$, we use $\mathcal {N}(h,t) \subseteq \mathcal {U}$ to denote the set of its associated users during time slot $t$.
To avoid the high overheads of re-association,
we assume that each user is associated with at most one AP in each time slot $t$, \textit{i.e.}, 
\begin{equation}  \label{constraint: number of users for each AP}
	\sum_{h \in \mathcal {H}} X_{uh}(t) \leq 1,
   \ \forall\,u \in \mathcal {U}.
\end{equation}
On the other hand, due to resource limit, each AP can associate with at most $M$ users in each time slot $t$, \textit{i.e.}, 
\begin{equation}  \label{constraint: number of associated APs}
	\sum_{u \in \mathcal {U}} X_{uh}(t) \leq M, \ \forall\,h \in \mathcal {H}.
\end{equation}

Regarding the cache placement (AP-file association), we model it as a bipartite graph $\widetilde{\mathcal {G}} \triangleq (\mathcal {H},\mathcal {F},\widetilde{\mathcal {E}})$ 
such that each edge $(h,f) \in \widetilde{\mathcal {E}}$ indicates that AP $h$ has files of type $f$ in its cache. 
Given that the cache placement is fixed, we use $\boldsymbol{Y}$ to denote the cache placement matrix with respect to $\widetilde{\mathcal {G}}$ such that $Y_{hf} =1$ if $(h,f) \in \widetilde{\mathcal {E}}$ and zero otherwise.

\subsection{The Transmission Model}
We assume that the wireless channels between users and APs are flat fading channels\cite{Tse2005fundamentals}. 
For each AP $h$, we assume its transmit power to be constant across time slots, denoted by $P_{h}$. Besides, we use $B_{h}(t)$ to denote the total bandwidth of AP $h$ and $B_{uh}(t)$ to denote the bandwidth allocated to user $u$ by AP $h$ within time slot $t$, such that\footnote{
		In practice, $B_{uh}(t)$ is usually lower bounded by some minimum bandwidth guarantee, \textit{e.g.}, $180$kHZ in the LTE standard\cite{Sesia2009lte}.
	}
\begin{equation}
	\sum_{u \in \mathcal {N}(h,t)}B_{uh}(t) = B_h(t).	
\end{equation}
Next, we define the maximum achievable rate of link $(u,h)$ during time slot $t$ as
\begin{equation}  \label{LinkCapacity}
	\begin{array}{l}
	C_{uh}(t) \triangleq \\
	\displaystyle
	\tau B_h(t)\cdot\mathbb{E}\bigg[\text{log}\left(1+\frac{P_hg_{uh}(t)|s_{uh}(t)|^2}{1+\sum_{h^{'} \in \mathcal {H}\backslash h}P_{h^{'}}g_{uh^{'}}(t)|s_{uh^{'}}(t)|^2} \right) \bigg],		
	\end{array}
\end{equation}
where $g_{uh}(t)$ is the large-scale fading gain due to path loss and shadowing,\footnote{
		With commonly adopted rate adaption schemes \cite{Biglieri1998fading}\cite{Ong2011ieee}, we assume that each AP $h$ is aware of the slowly varying path-loss coefficient $g_{uh}(t)$ for each user $u$.
	}
and $s_{uh}(t)$ is the small-scale fading gain that follows Rayleigh distribution.
Accordingly, we define the service rate over link $(u,h)$ within time slot $t$ as \cite{Bethanabhotla2015videoStreaming}
\begin{flalign}
  \nonumber  &\mu_{uh}(t) \triangleq \\
   & \tau B_{uh}(t)\cdot\mathbb{E}\Bigg[\text{log}\Bigg(1+\frac{P_hg_{uh}(t)|s_{uh}(t)|^2}{1+\sum_{h^{'} \in \mathcal {H}\backslash h}P_{h^{'}}g_{uh^{'}}(t)|s_{uh^{'}}(t)|^2} \Bigg) \Bigg].
   \label{userRate}
\end{flalign}
For each AP $h$, given a set of rate allocations $\{\mu_{uh}(t): u \in \mathcal {N}(h,t)\}$, they are said to be \textit{feasible} if and only if 
\begin{equation}\label{ChannelRateConstraint1}
    \sum_{u \in \mathcal {N}(h,t)}\frac{\mu_{uh}(t)}{C_{uh}(t)} = 1.
\end{equation}
For simplicity, we introduce variable $\nu_{uh}(t)$ for each user-AP pair $(u, h)$ such that $\nu_{uh}(t) = \frac{\mu_{uh}(t)}{C_{uh}(t)}$ if $X_{uh}(t) = 1$ (\textit{i.e.}, user $u$ is associated with AP $h$ within time slot $t$) and zero otherwise. 
We further define $\boldsymbol{\nu}(t) \triangleq \{\nu_{uh}(t) \vert u \in \mathcal {U}, h \in \mathcal {H}\}$ as the service rate allocation decisions made in time slot $t$.
Each entry $\nu_{uh}(t)$ can be viewed as the proportion of service rates allocated by AP $h$ to user $u$.
Accordingly, we can rewrite (\ref{ChannelRateConstraint1}) as the following equivalent constraint
\begin{equation}\label{ChannelRateConstraint2}
	\sum_{u \in \mathcal {U}}\nu_{uh}(t)X_{uh}(t) = 1, \ \forall h \in \mathcal {H}.
\end{equation}
Recall that each user can associate with at most one AP in one time slot. Therefore, within time slot $t$, the total service rate allocated to user $u$ is given by
\begin{equation}\label{ServiceRate}
    \mu_u(t) = \sum_{h \in \mathcal {H}}C_{uh}(t)\nu_{uh}(t)X_{uh}(t).
\end{equation}

\subsection{User Traffic Model}
During each time slot $t$, we assume that each user $u$ generates new requests for only one type of files.
Particularly, the probability of requesting each type $f$ of files is assumed subject to the Zipf distribution \cite{zink2009characteristics}.
Next, we define binary variable $I_{uf}(t)$ to indicate whether user $u$ requests files of type $f$ during time slot $t$. 
Accordingly, we have $\sum_{f \in \mathcal{F}} I_{uf}(t) = 1$ and for notational simplicity, we define $\boldsymbol{I}(t) \in \{0, 1\}^{U \times F}$ as the set of all such binary variables.
Meanwhile, we use $A_{u}(t)$ to denote the total size (in the units of Mbits) of files that are newly requested by user $u$ during time slot $t$,
which is assumed upper bounded by some constant $A_{\text{max}}$ and \textit{i.i.d.} across time slots with a finite expectation of $\mathbb{E}\{A_u(t)\}=\lambda_u$.
For notational simplicity, we define $\boldsymbol{A}(t) \triangleq \{A_{u}(t)I_{uf}(t), u \in \mathcal {U}, f \in \mathcal {F}\}$ as the vector of requested file sizes during time slot $t$.

\subsection{Predictive Scheduling Model}
We consider the case in which users' future request demands can be perfectly predicted in a finite lookahead time window.\footnote{Note that our model is not dependent on any particular request prediction techniques. In practice, such predictions can be achieved by exploiting various machine learning techniques \cite{NetflixPred}\cite{mao2017neural}.} 
With such predicted information, the upcoming requests can be pre-generated and appended to their corresponding file queues. 
Moreover, once given adequate service rates, users can pre-retrieve the files of such predicted requests before their actual arrivals, so that shorter delays and better quality of experience (QoE) can be achieved.
Such a mild assumption is reasonable, considering the wide adoption of effective prediction techniques in various scenarios (\textit{e.g.}, Netflix's preloading of videos to users based on its preference prediction\cite{NetflixPred}).

To formalize such a predictive scheduling mechanism, we adopt the \emph{lookahead time window} model in \cite{Huang2014When}.
Particularly, for each user $u$, we assume that its future request demands $\{A_u(t),\ldots,A_u(t+D_u-1)\}$ and $\{I_{uf}(t),\ldots,I_{uf}(t+D_u-1)\}$ are accessible by the user in a \emph{lookahead window} of size $D_{u}$ ($D_u \geq 1$). 
We use $\boldsymbol{D} \triangleq (D_1, D_2, \cdots, D_U)$ to denote the vector of prediction window sizes for all users. 
 
Note that file requests in the prediction windows can be pre-served before their actual arrivals. Therefore, the total size of unserved files for each time slot in prediction windows may decrease across time slots.
To record such a change, for each user $u$ and each time slot $t$, we use $\tilde{Q}_{uf}^{d}(t)$ to denote the total size of unserved files (of type $f$) that will be requested in time slot $(t+d)$, for $d \in \{ 0, 1, \dots, D_{u}-1 \}$.\footnote{In our model, we assume that new requests are appended to queues at the end of each time slot. Therefore, for the case with $d=0$, new requests are yet to be generated at the beginning of each time slot.}
Meanwhile, we use $\tilde{Q}_{uf}^{-1}(t)$ (with $d=-1$) to denote the total size of actually requested but unserved files (of type $f$) at the beginning of time slot $t$.
Then the total size of unserved files (including those requested or predicted but unserved) is given by 
$Q_{uf}(t) = \sum_{d=-1}^{D_{u}-1} \tilde{Q}^{d}_{uf}(t)$ for each user $u$. 

Accordingly, for each time slot $t$, 
we use $\tilde{\mu}_{uf}^{-1}(t)$ and $\{\tilde{\mu}_{uf}^{d}(t)\}_{d=0}^{D_u-1}$ to denote the service rates assigned to the requested but unserved files 
and the unserved files (both of type $f$) in the prediction window, respectively.
In practice, such service rates $\{\tilde{\mu}_{uf}^{d}(t)\}_{d=-1}^{D_u-1}$ can be assigned according to a certain discipline such as FIFO and LIFO. 
In this paper, we assume the adoption of the predictive scheduling policy that serves requests in each $Q_{uf}(t)$ in a FIFO and fully efficient manner\cite{Huang2014When}, such that
\begin{equation}\label{PredictiveRate}
   \sum_{f \in \mathcal {F}} \mu_{uf}(t) = \mu_{u}(t),   
\end{equation}
in which we define $\mu_{uf}(t) \triangleq \sum_{d=-1}^{D_{u}-1} \tilde{\mu}^{d}_{uf}(t)$ such that $\tilde{\mu}_{uf}^d(t) = 0$ if $Y_{hf} = 0$, and $\tilde{\mu}_{uf}^d(t) > 0$ only if $\tilde{\mu}_{uf}^{d-1}(t) \geq \tilde{Q}^{d-1}_{uf}(t)$. 
Intuitively, under such a policy, each user $u$ is ensured to utilize all of its allocated service rates to serve requests within its file queues $\{Q_{uf}(t)\}_{f \in \mathcal{F}}$ in the chronological order by their arrival times.

\begin{figure}[!t]
    \centering{
    \includegraphics[width=0.8\columnwidth]{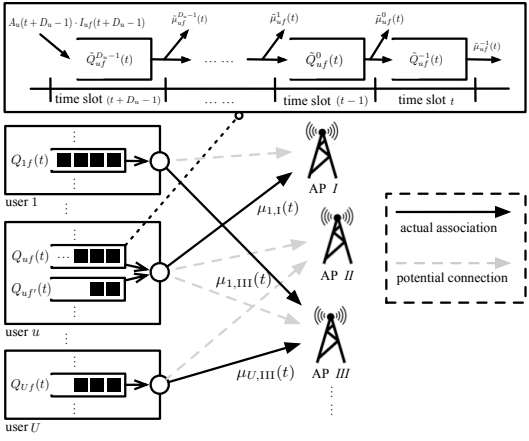}}
    \caption{Illustration of queueing dynamics in the system. 
    		During each time slot, each user has potential access (denoted by potential connections using dashed arrows) to a number of APs but can only send its requests to one of them. Each selected connection is called an actual association (denoted by a solid arrow).}
    	For example, there are three user-AP associations in the figure, including $(\text{user}\,1,\text{AP-III})$, $(\text{user}\,u,\text{AP-I})$, and $(\text{user}\,U,\text{AP-III})$. 
    	Besides, we also show how prediction queue backlogs $\{\tilde{Q}^{d}_{uf}(t)\}_{d=-1}^{D_u-1}$ of user $u$ and type $f$ are updated.
    \label{fig:Queueing}
 \end{figure}

By defining $[x]^+ \triangleq \text{max}\{x,0\}$, we can write the backlog update equations between consecutive time slots for each user $u$ and each file type $f$ as follows.\footnote{
		We set $\tilde{Q}_{uf}^{-1}(0) = 0$ and $\tilde{Q}_{uf}^d(0) = A_u(d)\cdot I_{uf}(d)$ for $0 \leq d \leq D_{u}-1$.
	}
\begin{enumerate}
  \item If $d = D_u-1$, then
  \begin{align}\label{queue1}
    \tilde{Q}_{uf}^{d}(t+1) = A_u(t+D_u)\cdot I_{uf}(t+D_u).
  \end{align}
  \item If $0 \leq d \leq D_u-2$, then
  \begin{align}
     \tilde{Q}_{uf}^{d}(t+1)=\bigg[ \tilde{Q}_{uf}^{d+1}(t)- \tilde{\mu}_{uf}^{d+1}(t) \bigg]^{+}. \label{queue2}
  \end{align}
  \item If $d = -1$, then
   \begin{align}
    \!\!\!\!\!\!\!\!\!\!\!\!\!\!\!\!\!\tilde{Q}_{uf}^{-1}(t+1)=\bigg[ \tilde{Q}_{uf}^{-1}(t)- \tilde{\mu}_{uf}^{-1}(t) \bigg]^{+}
       \!\!\!\!\!\!+  
       \bigg[\tilde{Q}_{uf}^{0}(t)- \tilde{\mu}_{uf}^{0}(t) \bigg]^{+}. \label{queue3}
  \end{align}
\end{enumerate}
We illustrate the above queueing dynamics in Figure \ref{fig:Queueing}.
 \setlength{\floatsep}{-10pt}


\subsection{System Objectives and Problem Formulation}
Given the settings of our model, we switch to specifying the system objectives for our problem formulation.
First, we introduce some definitions as follows. 
Particularly, we assume that each channel path-loss coefficient $g_{uh}(t)$ ($\forall (u,h)\!\in\!\mathcal {E}$) changes slowly during each time slot $t$. Then we define the random event occurred at the beginning of time slot $t$ and the scheduling policy as follows, respectively.

\textbf{Definition} $1$: The random event $\omega(t)$ occurred in time slot $t$ contains the slowly-varying channel path-loss coefficients, the amount of new request arrivals and corresponding requested file types. Therefore, we have
\begin{equation}\label{RandomEvent}
    \omega(t) = \{g_{uh}(t), A_u(t),I_{uf}(t),\ \forall u \in \mathcal {U},h \in \mathcal {H}, f \in \mathcal {F} \}
\end{equation}

\textbf{Definition} $2$: The scheduling policy $\{\alpha(t)\}_{t=0}^{\infty}$ is a sequence of control actions $\alpha(t)$ which comprises the user-AP association $\boldsymbol{X}(t)$ and bandwidth allocation $\boldsymbol{\nu}(t)$; \textit{i.e.},
\begin{align}
\alpha(t) = \{\boldsymbol{X}(t),\boldsymbol{\nu}(t)\}.	
\end{align}

\textbf{Definition} $3$: The feasible set of control actions $\mathcal {A}_{\omega(t)}$ in time slot $t$ includes all feasible control actions that satisfy constraints (\ref{constraint: number of associated APs}),(\ref{constraint: number of users for each AP}), and (\ref{ChannelRateConstraint2}) simultaneously; \textit{i.e.}, 
\begin{equation}
	\mathcal {A}_{\omega(t)} \triangleq \{ \alpha(t) \,\vert\, (\ref{constraint: number of users for each AP}),(\ref{constraint: number of associated APs}),(\ref{ChannelRateConstraint2}) \}.
\end{equation}

In our work, we aim to find a scheduling policy to jointly maximize the long-term time-average of total network throughput and achieve the stability of all queue backlogs in the system. They are defined as follows, respectively.

\textit{Network Throughput:}
We consider the following metric to characterize the throughput of each user $u$ in the system.
\begin{equation}\label{TimeAverageRate}
    \overline{\mu}_{u} \triangleq \lim_{t \rightarrow \infty}\frac{1}{t}\sum_{\tau = 0}^{t-1}\sum_{f \in \mathcal {F}}\mathbb{E}[\mu_{uf}(\alpha(\tau),\tau)],
\end{equation}
which is the time-average expectation of the total size of files retrieved by user $u$ in the long run.
For ease of notation, we define $\overline{\boldsymbol{\mu}} \triangleq (\overline{\mu}_1,\ldots,\overline{\mu}_U)$.
Accordingly, the total network throughput is given by $\phi(\overline{\boldsymbol{\mu}}) \triangleq \sum_{u \in \mathcal {U}}\overline{\mu}_u$. Since $\phi(\cdot)$ is a linear function, then we have $ \phi(\overline{\boldsymbol{\mu}}) = \overline{\phi(\boldsymbol{\mu})}$.

\textit{Queue Stability:} We define the long-term time-averaged expectation of the total queue backlog size in the system as
\begin{equation}\label{TimeAverageQueue}
    \overline{Q}_{uf} \triangleq \lim_{t \rightarrow \infty}\frac{1}{t}\sum_{\tau = 0}^{t-1}\mathbb{E}[Q_{uf}(\tau)]
\end{equation}
and adopt the notion of strong stability in \cite{neely2010stochastic}, \textit{i.e.},
\begin{equation}\label{stabilityConstraint}
    \overline{Q}_{uf} < \infty.
\end{equation}
Intuitively, constraint (\ref{stabilityConstraint}) ensures that each user's local queues will not be overloaded and their backlog sizes will not grow unboundedly.

With the above definitions, our problem formulation is given as follows.
\begin{eqnarray}
 \mathcal {P}1: \ 
	\max  && \overline{\phi(\boldsymbol{\mu})} \\
  	\text{Subject to} && \overline{Q}_{uf} < \infty \ \forall u \in \mathcal {U},f \in \mathcal{F} \label{QueueStabilityConstraint} \\
  &&\alpha(t) \in \mathcal {A}_{\omega(t)} \ \forall t.
\end{eqnarray}
Besides, we define $\phi^{\text{opt}}$ as the optimal throughput associated with the above problem $\mathcal {P}1$ and augmented with the rectangle constraint
    $\overline{\boldsymbol{\mu}} \in \mathcal {R}$,
where $\mathcal {R}$ is chosen large enough to contain a time-averaged throughput vector $\overline{\boldsymbol{\mu}}$ that is optimal to problem $\mathcal {P}1$.

\section{User-AP Association and Resource Allocation with Prediction}\label{PUARA}
By leveraging Lyapunov optimization techniques \cite{neely2010stochastic},
in this section, we show how we solve the problem $\mathcal {P}1$. 
Particularly, we demonstrate our devised algorithm called PUARA (Predictive User-AP Association and Resource Allocation) in Subsection \ref{PUARA}.1.
Then we conduct theoretical analysis to analyze its performances in Subsections \ref{PUARA}.2.

\subsection{Algorithm Design}\label{Algorithm}
We solve problem $\mathcal{P}1$ in an \textit{incremental} way.
First, we consider a special case in which the system proceeds \textit{without prediction}, 
\textit{i.e.}, $D_u = 0$ for each user $u \in \mathcal{U}$.
In such a case, we adopt Lyapunov optimization techniques \cite{neely2010stochastic} to solve problem $\mathcal {P}1$.
By applying such techniques, we decouple problem $\mathcal{P}1$ into a series of sub-problems over time slots. 
Specifically, in each time slot $t$, we aim to solve the following optimization problem.
\begin{align}
 \mathcal {P}2:  \text{min} & \ \  -V\sum_{u \in \mathcal {U}}\sum_{f \in \mathcal {F}}\mu_{uf}(t) - \sum_{u \in \mathcal {U}}\sum_{f \in \mathcal {F}}Q_{uf}(t)\mu_{uf}(t) \label{drift-plus-penalty} \\
  \nonumber  \text{Subject to} & \ \ (\ref{constraint: number of users for each AP}),(\ref{constraint: number of associated APs}),(\ref{ChannelRateConstraint2}),
\end{align}
where parameter $V$ is a positive constant.
After rearranging the objective function of problem $\mathcal {P}2$, we can rewrite it as
\begin{eqnarray}
    -\sum_{u \in \mathcal {U}}\sum_{f \in \mathcal {F}}\Big[V + Q_{uf}(t) \Big]\mu_{uf}(t) \label{drift-plus-penalty2}.
\end{eqnarray}
Next, we define 
$\mathcal {M}_{uh}(t) \triangleq C_{uh}(t) \cdot \big[\sum_{f \in \mathcal{F}} (V + Q_{uf}(t)) Y_{hf}\big]$.
Note that $\mathcal {M}_{uh}(t)$ is a constant within the current time slot $t$ since all queue backlog sizes $Q_{uf}(t),\forall u \in \mathcal {U}, f \in \mathcal {F}$ are given at the beginning of time slot $t$.
Then problem $\mathcal{P}2$ is equivalent to the following problem. 
\begin{eqnarray}\label{sub_objectiion}
    \nonumber \mathcal {P}_{in}: \max\limits_{\boldsymbol{\nu}(t),\boldsymbol{X}(t)} && \sum_{h \in \mathcal {H}}\sum_{u \in \mathcal {U}}\mathcal {M}_{uh}(t)\nu_{uh}(t) X_{uh}(t) \\
    && \\
   \nonumber \text{Subject to} && (\ref{constraint: number of users for each AP}),(\ref{constraint: number of associated APs}),(\ref{ChannelRateConstraint2}). 
\end{eqnarray}
Problem $\mathcal {P}_{in}$ is an integer programming problem with two types of decision variables coupled in product form: 
user-AP association $\boldsymbol{X}(t)$ and bandwidth allocation $\boldsymbol{\nu}(t)$. 
Such problems are in general $\mathcal{NP}$-hard to solve.
However, by observation, we find that given a fixed value for each entry in set $\boldsymbol{X}(t)$, problem $\mathcal{P}_{in}$ is actually a linear programming problem with respect to variables $\boldsymbol{\nu}(t)$, which aims to find a set of weights $\mathcal{M}_{uh}(t)$ with a maximum sum. For such a problem, its solutions can be always found at the boundary of its domain. 
Moreover, recall that each variable $\nu_{uh}(t)$ is defined as $\frac{\mu_{uh}(t)}{C_{uh}(t)}$ if $X_{uh}(t) = 1$ and zero otherwise.
Therefore, we can rewrite problem $\mathcal {P}_{in}$ as follows.
\begin{eqnarray}\label{sub_objectiion}
     \mathcal {P}_{in}': \max\limits_{\boldsymbol{X}(t)} && \sum_{h \in \mathcal {H}}\sum_{u \in \mathcal {U}}\mathcal {M}_{uh}(t)X_{uh}(t) \label {P_in2}\\
    \nonumber \text{Subject to} &&  
    \boldsymbol{X}(t) \in \{0, 1\}^{U \times H},
    \\
   \nonumber && (\ref{constraint: number of users for each AP}),(\ref{constraint: number of associated APs}).
\end{eqnarray} 
The equivalence relationship between problems $\mathcal {P}_{in}$ and $\mathcal {P}_{in}'$ is formalized by the following lemma, for which the proof is relegated to Appendix A.
\begin{lemma}
	\textit{Problems $\mathcal {P}_{in}$ and $\mathcal {P}_{in}'$ are equivalent.}
\end{lemma} 

We find that problem $\mathcal {P}_{in}'$ is actually equivalent to a modular maximization problem over two matroid constraints.\footnote{We relegate the proof of the matroid structure of problem $\mathcal{P}'_{in}$ to Appendix \ref{matroid-transformation}. For more details about matroids and modular functions, please refer to \cite{edmonds1971matroids}.}
Such a problem structure allows us to design computationally efficient algorithms with provable performance guarantee. 
Note that
for general modular maximization problems subject to $p$ matroid constraints\cite{calinescu2007maximizing}, 
greedy algorithms have been proven as effective approaches with a tight approximation ratio of ${1}/{p}$\cite{edmonds1971matroids,jenkyns1976efficacy,korte1978analysis}.
\begin{algorithm}[!h]
        \caption{Greedy User-AP Association algorithm}
        \begin{algorithmic}[1]  \label{greedy alg.}
              \STATE At beginning of time slot $t$, APs collect instant maximum achievable rates $C_{uh}(t)$ and queue backlog sizes $Q_{uf}(t)$, $\forall\,u,h,f$, and upload them to the operation center.
              \STATE The operation center sets $\mathcal{X}  \leftarrow \emptyset$, 
              			 $\mathcal {W} \leftarrow \{ \mathcal{X}_{u}^{h} \}_{u,h}$, then: \\
              \STATE \textbf{while} $|\mathcal {W}| > 0$  \textbf{do}
              \STATE\ \ \ \  
              $\mathcal{X}_{u^{*}}^{h^{*}} \in \mathop{\text{argmax}}\limits_{\mbox{\tiny
              	  \!\!\!$
              	\begin{array}{c}
	              {\mathcal{X}_{u}^{h} \in \mathcal {V}},\\
    	          {\{ \mathcal{X} \cup \mathcal{X}_{u}^{h} \} \subseteq \mathcal {C}}
    	        \end{array}
    	          $\!\!\!}}  \mathcal{M}_{uh}(t)$.
              \STATE \ \ \ \ Update $\mathcal {W} \leftarrow \mathcal {W} \backslash \{ \mathcal{X}_{u^{*}}^{h^{*}} \}$.
              \STATE \ \ \ \ Update $\mathcal{X} \leftarrow \mathcal{X} \cup \{ \mathcal{X}_{u^{*}}^{h^{*}} \}$.
              \STATE  \textbf{end while}
              \STATE The operation center spreads the decision $\boldsymbol{X}$ to APs, where ${X}_{uh} = 1$ if $\mathcal{X}_{u}^{h} \in \mathcal{X}$ and zero otherwise.
        \end{algorithmic}
\end{algorithm}

Accordingly, we devise an efficient user-AP association algorithm to solve problem $\mathcal{P}'_{in}$ in a greedy manner.
Its pseudocode is shown in Algorithm \ref{greedy alg.}, in which we define $\mathcal {C}$ as the intersection of the feasibility regions for constraints (\ref{constraint: number of users for each AP}) and (\ref{constraint: number of associated APs}),
 $\mathcal{X}$ as the set of activated user-AP associations, and $\mathcal {W}$ as the set of inactivated association pairs. 

In Algorithm \ref{greedy alg.} (line $2$),
initially, we set $\mathcal{X} = \emptyset$ and $\mathcal {W} = \{ \mathcal{X}_{u}^{h} \vert u\in \mathcal {U},h \in \mathcal {H}\}$. 
During each iteration (lines $4$ -- $6$), by Algorithm \ref{greedy alg.}, the inactivated association $\mathcal{X}_{u}^{h} \in \mathcal {W}$ with the highest marginal value $\mathcal{M}_{uh}(t)$ is selected subject to constraints (\ref{constraint: number of users for each AP}) and (\ref{constraint: number of associated APs}), then added to the set $\mathcal{X}$. 
The association pair $\mathcal{X}_{u^{*}}^{h^{*}}$ is then removed from $\mathcal{W}$. 
In addition, all $\mathcal{X}_{u^{*}}^{h}$ for $h \in \mathcal{H}\backslash\{ h^{*} \}$ are also removed since each user are assumed to associate with at most one AP.
Such a procedure terminates when the set $\mathcal{W}$ becomes empty.

Next, we consider more general cases \textit{with prediction}.  
By integrating predictive scheduling with Algorithm 1, we propose PUARA, a predictive user-AP association scheme which exploits predicted information to solve the following problem during each time slot $t$.
\begin{eqnarray}
 \nonumber \mathcal {P}3:  \text{min} && -V\sum_{u \in \mathcal{U}}\sum_{f \in \mathcal{F}}\sum_{d = -1}^{D_u - 1}\tilde{\mu}_{uf}^d(t) \\
  && - \sum_{u \in \mathcal {U}}\sum_{f \in \mathcal {F}}
  		Q_{uf}(t)
  		\sum_{d = -1}^{D_u - 1} \tilde{\mu}_{uf}^d(t) \label{PredictiveMinimizedrift-plus-penaltyProblem} \\
 \nonumber \text{Subject to} && (\ref{constraint: number of users for each AP}),(\ref{constraint: number of associated APs}),(\ref{ChannelRateConstraint2}).
\end{eqnarray}
We show the pseudocode of PUARA in Algorithm \ref{predictive alg.}.

\begin{algorithm}[!h]
        \caption{Predictive User-AP Association and Resource Allocation (PUARA) algorithm}
        \begin{algorithmic}[1] \label{predictive alg.}
              \STATE Initialize $t \leftarrow 0$, $\boldsymbol{Q}(0) \leftarrow \boldsymbol{0}$.
              \STATE \textbf{Repeat} \textbf{do}
              \STATE \ \ APs collect instant queue lengths $\{Q_{uf}(t)\}_{f \in \mathcal{F}}$ from \\ \ \ each user $u$, and upload such information with instant \\ \ \  rates $\{C_{uh}(t)\}_{u, h}$ to the operation center.
              \STATE \ \ The operation center runs Algorithm \ref{greedy alg.} to obtain $\boldsymbol{X}(t)$.
              \\ \ \ and spreads the decision $\boldsymbol{X}(t)$ to all APs.
              \STATE \ \ For each user-AP pair $(u, h)$ such that $X_{uh}(t)=1$, AP \\ \ \  $h$ allocates service rate $\mu_{u}(t) = C_{uh}(t)$ to user $u$.
              \STATE \ \ Given the service rate $\mu_{u}(t)$, each user $u$ downloads its \\ \ \ 
              	 requested files of type $f$ with $Q_{uf}(t) = \underset{f'\in \mathcal{F}}{\max} Q_{uf'}(t)$ \\ \ \ 
              	 from its associated AP $h$ in a FIFO and fully efficient \\ \ \ manner.
              \STATE \ \ Each user $u$ updates $\{Q_{uf}(t+1)\}_{f \in \mathcal{F}}$ by (\ref{queue1})-(\ref{queue3});
              \STATE \ \ t $\leftarrow$ t + 1.
              \STATE \textbf{end}
        \end{algorithmic}
\end{algorithm}

\textbf{Remark 1:}
Recall that we define $\mathcal {M}_{uh}(t) \triangleq C_{uh}(t) \cdot \big[\sum_{f \in \mathcal{F}} (V + Q_{uf}(t)) Y_{hf}\big]$. 
In Algorithm \ref{greedy alg.}, the parameter $V$ actually plays a central role in controlling the balance between network throughput maximization and queue stability. 
To demonstrate such an insight, we first note that the value of $\mathcal {M}_{uh}(t)$ for each user-AP pair $(u, h)$ is a constant within each time slot $t$. 
Then we can see that when the value of parameter $V$ is sufficiently large ($V \gg Q_{uf}(t)$), for each AP $h$, the difference of maximum achievable rates among its candidate users would be more significant than the difference among their queue backlogs. As a result, AP $h$ will be more willing to associate those users with greater maximum achievable rates, which conduces to a higher throughput. In contrast, when the value of parameter $V$ is small, then for each AP $h$, the difference among users' queue backlog sizes (the total size of each user's unserved files) would be more significant. Therefore, AP $h$ will be more willing to associate those users with more unserved files, which conduces to stabilizing queue backlogs in the system. In practice, the value of parameter $V$ can be chosen based on the design objective of real systems.

\textbf{Remark 2:}
Under PUARA, during each time slot $t$, each user $u$ will utilize all of its instant service rate $\mu_u(t)$ (allocated by its associated AP) to download its requested files of type $f$ such that $Q_uf(t) = \arg\max_{f'\in \mathcal{F}} Q_{uf'}(t)$ (if there are more than one file queues with the maximum size, then spread the service rate $\mu_u(t)$ evenly among them). Specifically, for each file queue $Q_{uf}(t)$, its allocated service rates will be first utilized to download the files that are actually requested by time slot $t$. If all such files have been downloaded and there are surplus service rates, then the user will utilize them to serve predicted requests in $Q_{uf}(t)$ in a chronological order by their predicted arrival times until all service rates are depleted. Such pre-service is applicable in practice. For example, Netflix preloads videos to users based on their predicted preferences, in which user requests can be pre-generated and pre-served\cite{NetflixPred}.

\subsection{Performance Analysis of PUARA}\label{JUARAComplexity}

{\indent \textit{Computational Complexity:} During each time slot, the computational complexity of PUARA mainly lies in the greedy AP-user association procedure in Algorithm 1. Accordingly, the computational complexity of the greedy algorithm is
$    \mathcal {O}(U T_m)$,
where $T_m$ denotes the computation complexity for searching for the element $\mathcal{X}_{u^\star}^{h^\star}$. 
Specifically, by Algorithm 1, the procedure begins with an empty set. 
During each iteration, it adds one element with the highest marginal value to the set while maintaining the feasibility of the solution. Since the objective function is modular, the marginal value of the elements decreases as we add more elements to set $\mathcal{X}$.
When the largest marginal value is zero by some iteration, the procedure should stop. 
Since each user is assumed to associate with at most one AP, then at most $U$ iterations will be taken. Each iteration involves evaluating the marginal value of at most $U \times H$ elements. 
Accordingly, the computation complexity for searching for element $\mathcal{X}_{u^\star}^{h^\star}$ is $\mathcal{O}(T_m) = \mathcal{O}(UH)$. As a result, the overall computation complexity of PUARA is $\mathcal {O}(U^2H)$.}

\textit{Optimality:} 
To characterize the throughput incurred by PUARA compared to the optimal value of $\mathcal{P}1$, 
we refer to its scheduling as \emph{imperfect scheduling}\cite{ImperfectSchedule2006TON}\cite{ResourceAllocationCLC2006georgiadis}.
Then during each time slot $t$, the resulting service rates $\boldsymbol{\mu}(t) \in \mathcal {R}$ satisfy
\begin{eqnarray}
 \nonumber  && \sum_{u \in \mathcal {U}}\sum_{f \in \mathcal {F}}\Big[V + Q_{uf}(t) \Big]\mu_{uf}(t) \\
   \nonumber &\geq& \beta\mathop{\text{max}}\limits_{\boldsymbol{\mu}(t) \in \mathcal {R}}\Big\{\sum_{u \in \mathcal {U}}\sum_{f \in \mathcal {F}}\Big[V + Q_{uf}(t) \Big]\mu_{uf}(t)\Big\},\\
   \label{GreedyGap1}
\end{eqnarray}
where we can view constant $\beta \in (0, 1]$ as the approximation ratio of PUARA. Note that when $\beta = 1$, it reduces to the optimal scheduling for problems $\mathcal{P}_{2}$ and $\mathcal {P}_{in}$. 
With such a notion, we have $\beta=1/2$ for PUARA, which means that it solves problem $\mathcal{P}'_{in}$ with an approximation of ${1}/{2}$. The proof is relegated to Appendix \ref{imperfect scheduling}.

\textit{Impacts on Throughput and Delay:} 
We define $\phi_{av}^{\text{PUARA}}$ and $Q_{av}^{\text{PUARA}}$ as the long-term time-averages of the expected network throughput and the expected total queue backlog size incurred by PUARA, respectively.
Then the performance of PUARA without prediction can be characterized by the following theorem.
\begin{theorem}\label{Theorem1}
	\textit{
		Without prediction, \textit{i.e.}, given $D_{u}=0$ for $u\in\mathcal{U}$, we have the following upper bound on the long-term time-average of overall network throughput under PUARA.
\begin{equation}
	\begin{array}{rl}  \label{Theorem2NetworkUtility}
		  \phi_{av}^{\text{PUARA}} \triangleq & 
  \displaystyle
  \liminf_{t \rightarrow \infty} \phi(\frac{1}{t} \sum_{\tau = 0}^{t-1} \mathbb{E}\{\boldsymbol{\mu}(\tau)\}) \\
  \geq & \displaystyle
  \beta \phi^{\text{opt}} - \frac{\mathcal {K}}{V}.
	\end{array}	
\end{equation}
Besides, the upper bound on the corresponding long-term time-average of the total queue backlog size is given by
\begin{equation}
	\begin{array}{rl}
  Q_{av}^{\text{PUARA}} \triangleq & 
  \displaystyle
  \limsup_{t \rightarrow \infty} \frac{1}{t} \sum_{\tau = 0}^{t-1}\sum_{u}\sum_{f}\mathbb{E}\{Q^{\text{sum}}_{uf}(t)\} \\
 \leq & \displaystyle \frac{\mathcal {K} + V(\phi^{\text{max}}-\phi_{\theta})}{
 \beta\theta},		
	\end{array}
\label{Theorem2Queue}
\end{equation}
where $\mathcal {K} = \frac{U}{2}(\mu_{\text{max}}^2 + A_{\text{max}}^2)$.
}
\end{theorem}
The proof is relegated to Appendix \ref{appdix8}.

\textbf{Remark:}
Theorem 1 also implies that, without prediction, PUARA achieves a tunable $[O(1/V), O(V)]$ tradeoff with respect to parameter $V$ 
between the time-average network throughput and the time-average total queue backlog size.
In general, a large value of $V$ encourages PUARA to distribute more user requests to those APs with greater service capacities so as to maximize the overall network throughput. 
However, this may also result in the overloading of such APs and hence a increased total queue backlog size. 
In contrast, a small value of $V$ will incur a more even distribution of user requests among APs. 
The price is that some requests may be delivered over wireless channels with poor conditions, thereby leading to a degraded throughput. 
In practice, the choice of the value of parameter $V$ depends on the particular objectives of system performance metrics.

Moreover, with prediction, PUARA can further break the $[O(1/V), O(V)]$ performance barrier with a notable reduction in the average queue backlog size of the system.
To characterize such benefits, we adopt the proof techniques in \cite{Huanglongbo2011TAC} to show that the queue vector in the system is within $O(\text{log}(V))$ distance away from a fixed point. 
Particularly, we have the following theorem.
\begin{theorem}
\textit{
Given predicted information, if FIFO queueing discipline is adopted and $D_u = O\big(\frac{1}{A_{\text{max}}}[q_u^* - G -K(\text{log}(V))^2 - \mu_{\text{max}}]^+ \big)$ for each user $u \in \mathcal {U}$, then compared to the non-prediction case, PUARA can achieve an average queue backlog size reduction by at most $\sum_{u \in \mathcal {U}} D_u\big[\lambda_u - O(\frac{1}{V^{\log(V)}}) \big]^+$.}
\end{theorem}
The proof is relegated to Appendix \ref{Theorem-2-proof}.
Theorem 2 shows that with predictive scheduling, the time-average total queue backlog size reduction is roughly proportional to the value of $\sum_{u \in \mathcal {U}}\lambda_u D_u$. 
Such a result implies that more predicted information (increasing prediction window sizes) conduces to shortening the total queue backlog size in the system, which implies a shorter average delay by \textit{Little's law}\cite{little1961proof}.

\section{Simulation Results}\label{Simulation}
In this section, we conduct simulations to evaluate the performance of PUARA under various settings. 
In the following subsections, we first demonstrate our simulation settings in Section \ref{subsec: basic settings} (the key parameter settings are summarized in Table \ref{tab:Parameter settings}). 
Then we present and discuss our simulation results under perfect and imperfect prediction in Sections \ref{subsec: sim results under perfect prediction} and \ref{subsec: sim results under imperfect prediction}, respectively.

\begin{table}[!t]
    \caption{Simulation settings}
    \vspace{-0.4cm}
    \begin{center}
        \begin{tabular}{lll}
          \hline\hline
          \textbf{Parameter} & \textbf{Description} & \textbf{Setting} \\
          \hline\hline
          $H$ & Number of APs & $9$ \\
          \hline
          $U$ & Number of users & $100$ \\
          \hline
          \multirow{3}*{$M$} & Maximum number of users asso-  & \multirow{3}*{$12$} \\
          & ciated with each AP during each & \\ 
          & time slot & \\
          \hline
          $\xi$ & Minimum bandwidth ratio for APs & $0.05$ \\
          \hline
          $P$ & Transmitting power for each AP & $10^8$  \\
          \hline
          $F$ & Total number of file types & $4$ \\
          \hline
          \multirow{2}*{$N$} & Number of types of files that & \multirow{2}*{$3$} \\
          & can be cached by each AP & \\
          \hline
          \multirow{2}*{$n_r$} & Parameter of Zipf distribution for &
          \multirow{2}*{$0.56$} \\
          & the types of requested files & \\
          \hline
          \multirow{2}*{$D$} & 
          	Size of the prediction lookahead   & \multirow{2}*{$\{0, 1, \dots, 60\}$} \\
          	& window for each user & \\
          \hline
          	$e_{\text{type}}$ & File type prediction error rate & $[0.0, 0.4]$ \\
          \hline
          	$e_{\text{size}}$ & File size prediction error rate & $[0.0, 0.4]$ \\
          \hline
          	$V$ & Value of control parameter & $[1, 10^4]$ \\
          \hline
          $f_{0}$ & Carrier frequency & $2.4$GHz \\
          \hline
        \end{tabular}
    \end{center}
    \label{tab:Parameter settings}
\end{table}

\subsection{Basic Settings}  \label{subsec: basic settings}
We consider a wireless caching network within an area of $50 \times 50$$m^2$, in which there are $9$ APs and $100$ randomly uniformly distributed users. 
Each AP is associated with at most $12$ users, given
the minimum bandwidth ratio $\xi$ as $0.05$ and $18$-MHz bandwidth for operation. 
The transmit power of each AP is fixed as $P = 10^8$. 
Based on the WINNER II channel model under small-cell scenario\cite{khan2011winner}, 
we set the path-loss coefficients for user-AP pair $(u, h)$ as
\begin{equation}\label{LossCoeff}
    g_{uh}(t) = 10^{-\frac{\text{PL}(d_{uh}(t))}{10}},
\end{equation}
where $d_{uh}$ denotes the distance from user $u$ to AP $h$ during time slot $t$,
and function $\text{PL}(\cdot)$ is defined as
\begin{equation}\label{PL}
    PL(d) \triangleq C_1 \log(d) + C_2 + C_3 \log(f_0/5) + \mathcal {X}_{dB},
\end{equation}
in which the carrier frequency $f_0=2.4$GHz, the shadowing log-normal variable $\mathcal {X}_{dB}$ has a variance of $\sigma_{dB}^2$, and coefficients $C_1$, $C_2$, $C_3$ denote specific constants under different communication conditions.
In our simulations, the communication over each link is under either line-of-sight (LOS) and non-line-of-sight (NLOS) condition independently with probability $p_l(d)$ and $1-p_l(d)$, respectively, such that
\begin{equation}\label{pl(d)}
    p_l(d) \triangleq \left\{
    \begin{array}{lc}
    1, & \text{if } d\leq 3m,  \\
    1-0.9\left[
    	1-(1.24-0.6\text{log}(d))^3
    \right]^{1/3}, & \text{otherwise.}
    \end{array}
    \right.
\end{equation}
Under line-of-sight (LOS) condition, we set $C_1 = 18.7$, $C_2 = 46.8$, $C_3 = 20$, and $\sigma_{dB}^2 = 9$;
under non-line-of-sight (NLOS) condition, $C_1 = 36.8$, $C_2 = 43.8$, $C_3 = 20$, and $\sigma_{dB}^2 = 16$. 
The total number of file types requested by users is $4$, while each AP is only able to cache $3$ of them. 
For each user, we generate its requests for files of type $f \in \mathcal {F}$ according to Zipf distribution \cite{ji2016wirelessD2D,zink2009characteristics,Golrezaei2014scaling}, with probability $p_f = \frac{f^{-\eta_r}}{\sum_{i \in \mathcal {F}}i^{-\eta_r}}$,
where $\eta_r=0.56$. 
We assume that all requests are served in a \textit{first-in-first-out} (FIFO) and fully efficient manner.
In addition, we vary the lookahead window size $D$ from $0$ to $60$ for all users, and the value of parameter $V$ from $1$ to $10^4$. 
We run the simulation for each combination of settings over $5\times 10^{5}$ time slots. Each time slot has a length of $10$ms.

\begin{figure}[!t]
   \begin{minipage}[t]{1.0\linewidth}
        \centering
        \subfigure[Time-averaged Throughput vs. $V$]{\includegraphics[width=0.7\columnwidth]{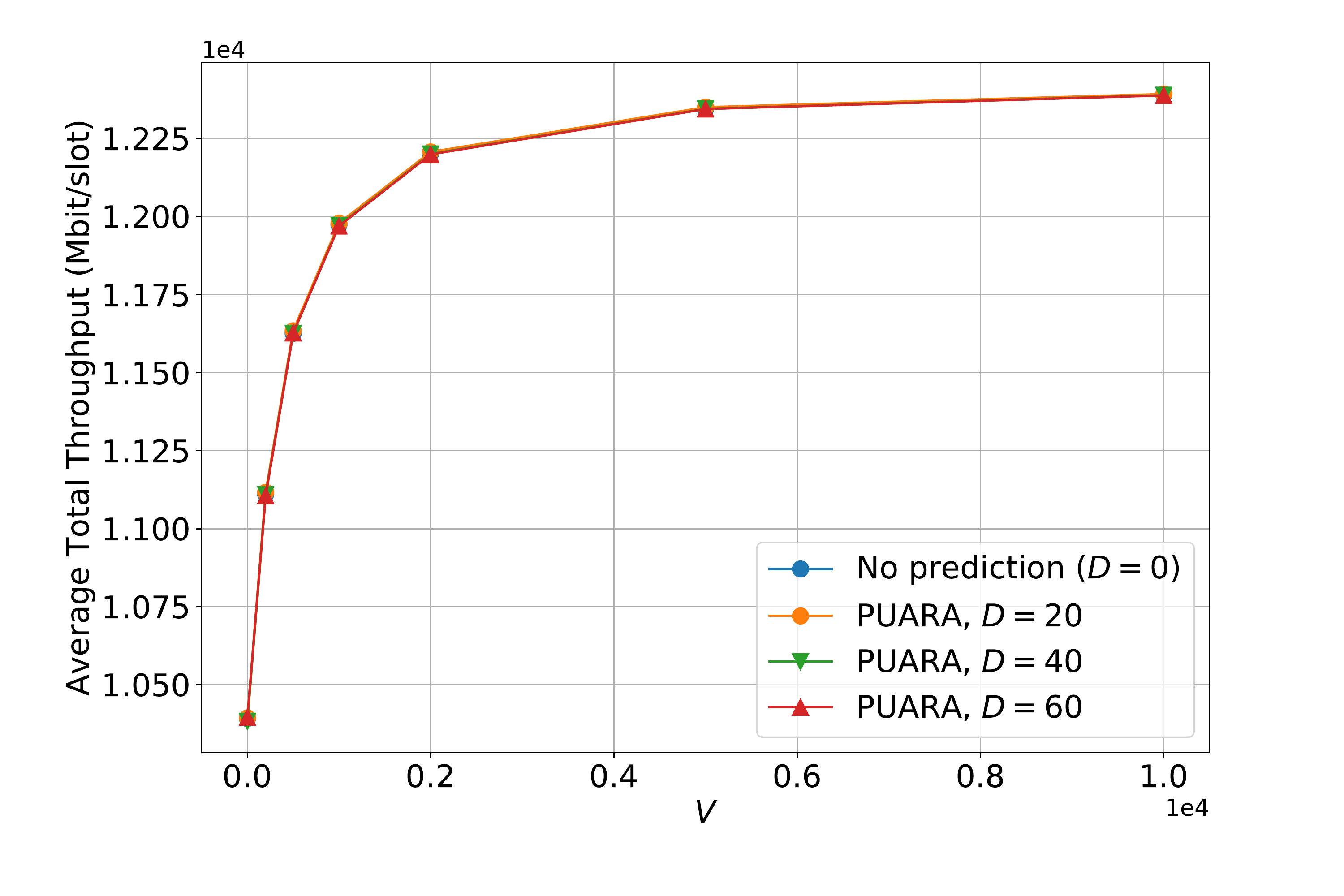}}\label{fig:AvgThroughputDiffD}
        \hspace{1cm}
        \subfigure[Average Delay of the System vs. $V$]{\includegraphics[width=0.7\columnwidth]{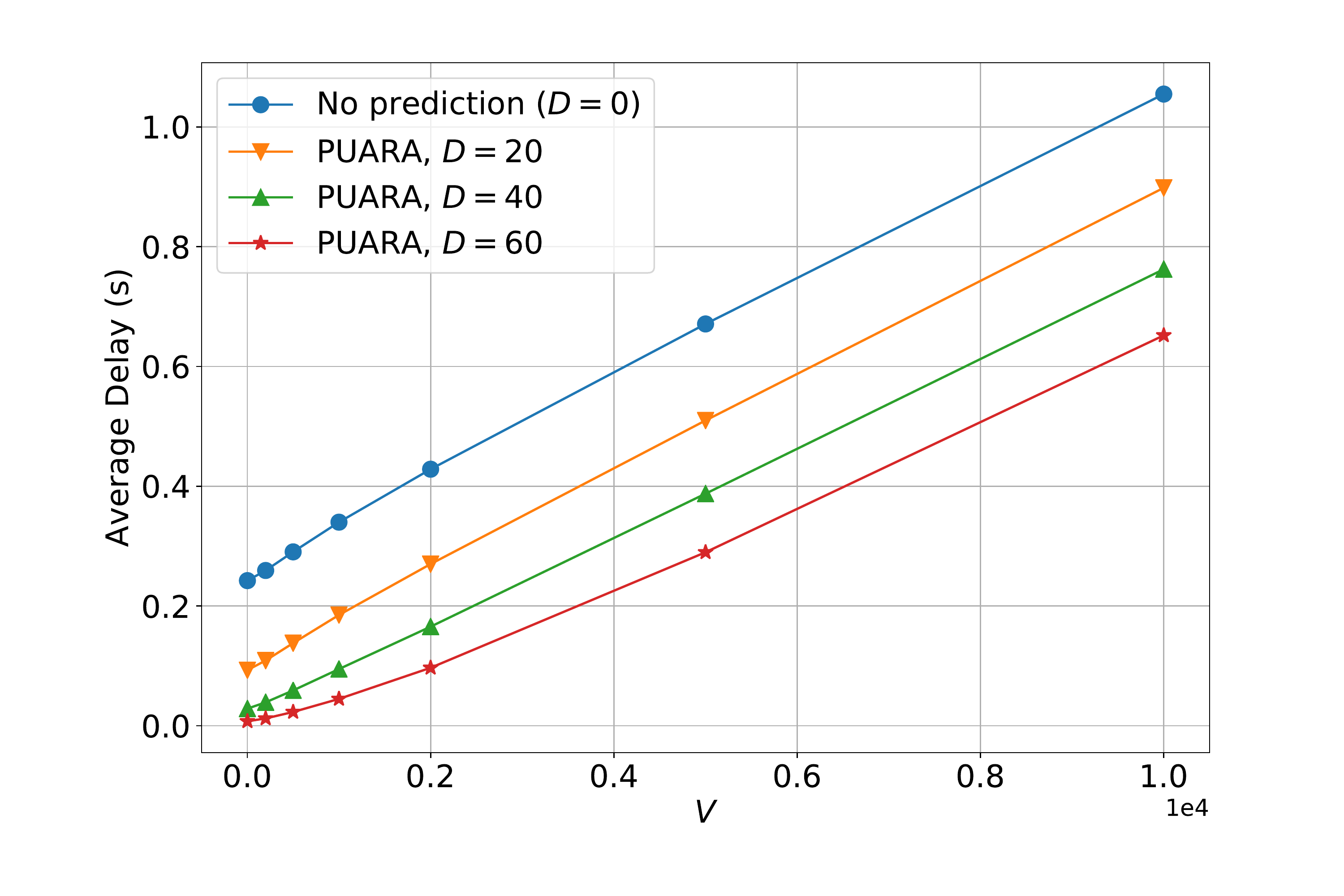}}\label{fig:AvgActualQueueDiffD}
   \end{minipage}
   \caption{
   	Performance of PUARA under different prediction window sizes given the maximum arrival rate $A_{\text{max}} = 100\text{Mbits}$. 
       As the value of parameter $V$ increases, the average network throughput approaches the optimum at the cost of an increased average delay for the system. 
       With more future information (as window size $D$ increases from $0$ to $60$), the average delay reduction is at least $38.2\%$ (when $V=10^4$).
   }
   \label{fig:JUARA_PUARAComp}
\end{figure}
\setlength{\textfloatsep}{2pt}

\subsection{Performances under perfect prediction}  \label{subsec: sim results under perfect prediction}
In this subsection, we evaluate the performance of PUARA under perfect prediction.

\textbf{Performance with different prediction window sizes:}
In Figure \ref{fig:JUARA_PUARAComp}, we show the performance of PUARA under different prediction window sizes (also the non-prediction case with $D=0$) with the maximum network traffic rate $A_{\text{max}}=100\,\textit{Mbits}$. 
Particularly, from Figure \ref{fig:JUARA_PUARAComp}(a), we see that in general, the time-averaged network throughput ascends and gradually flattens as the value of parameter $V$ increases from $1$ to $100$. 
Moreover, we find that increasing the value of prediction window size $D$ has a negligible impact on the time-averaged network throughput. 
This is because given that the total size of requested over a fixed period of time is a constant, predictive scheduling merely advances the service of part of files and generally makes no improvement in the throughput. 
Nonetheless, the pre-service of files conduces to the reduction in the average delay of the system.\footnote{
		In our simulations, the average delay of the system is defined as $\sum_{u \in \mathcal{U}} \sum_{f\in\mathcal{F}} \bar{Q}_{uf} / (\sum_{u \in \mathcal{U}} \bar{A}_{u})$. 
		Note that $\bar{Q}_{uf}$ denotes the time-averaged backlog size of user $u$'s queue with respect to file type $f$ and $\bar{A}_{u}$ denotes the time-averaged size of files requested by each user $u$. By Little's theorem \cite{little1961proof}, this is equal to the average delay for transmitting each Mbits of data in the system.}
We show the results in Figure \ref{fig:JUARA_PUARAComp} (b), in which the average delay of the system is reduced by at least $38.2 \%$ (when $V=10^4$) as the window size $D$ increases to $60$.
Such results also verify our theoretical analysis in Theorems 1 and 2, with respect to the $[O(1/V), O(V)]$ throughput-delay tradeoff (with $D=0$) and the backlog reduction (with $D>0$), respectively.

\begin{figure}[!t]
   \begin{minipage}[t]{1.0\linewidth}
        \centering
        \subfigure[Throughput-delay curve under different values of maximum arrival rate $A_{\text{max}}$ with the user number $U = 100$ and prediction window size $D =10$.]{\includegraphics[width=0.75\columnwidth]{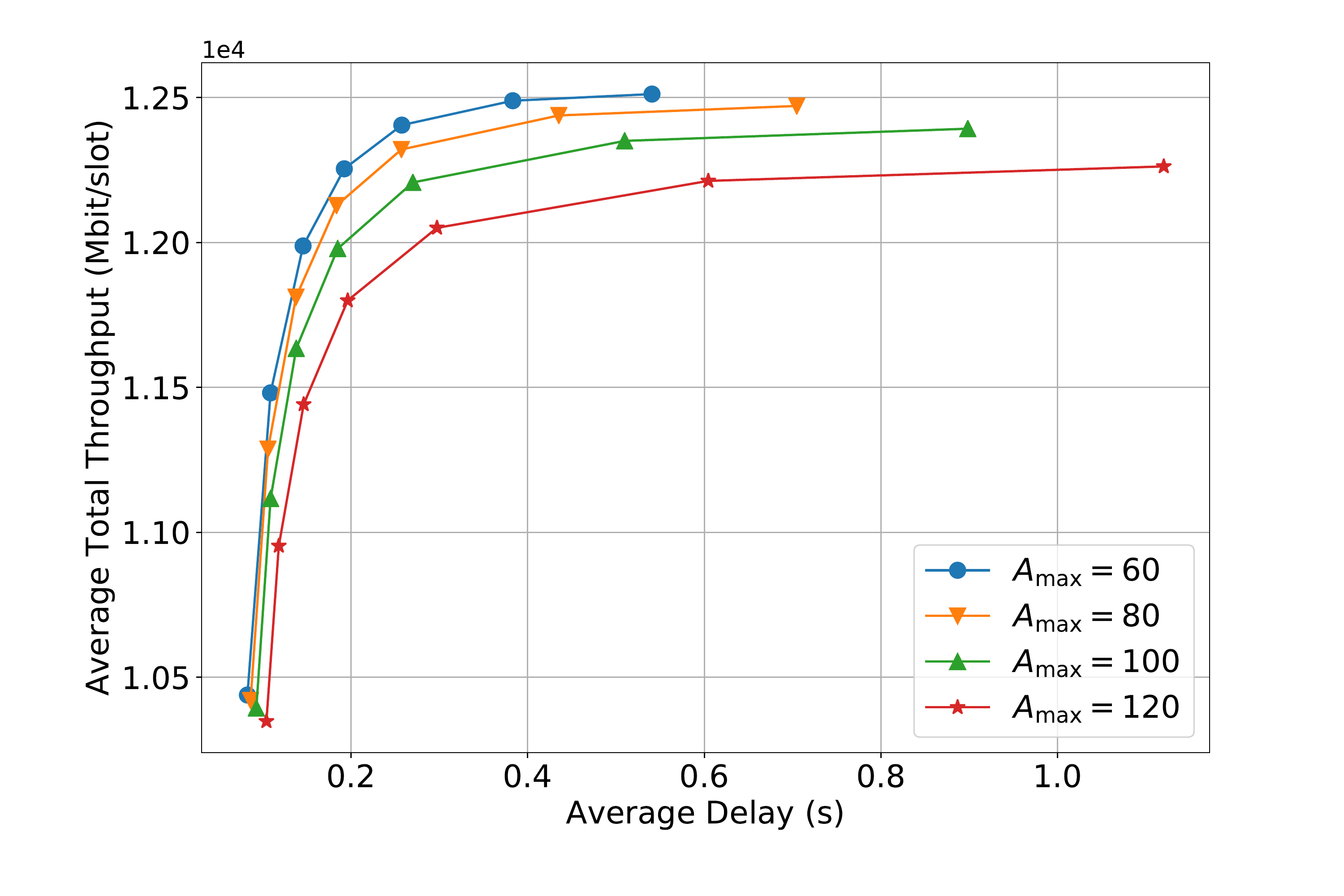}}\label{fig:AvgThroughputDiffworkload}
        \hspace{1cm}
        \subfigure[Average delay of the system against average throughput per user under different values of user number $U$, with the maximum arrival rate $A_{\text{max}}=100\text{Mbits}$ and prediction window size $D = 10$.]{\includegraphics[width=0.75\columnwidth]{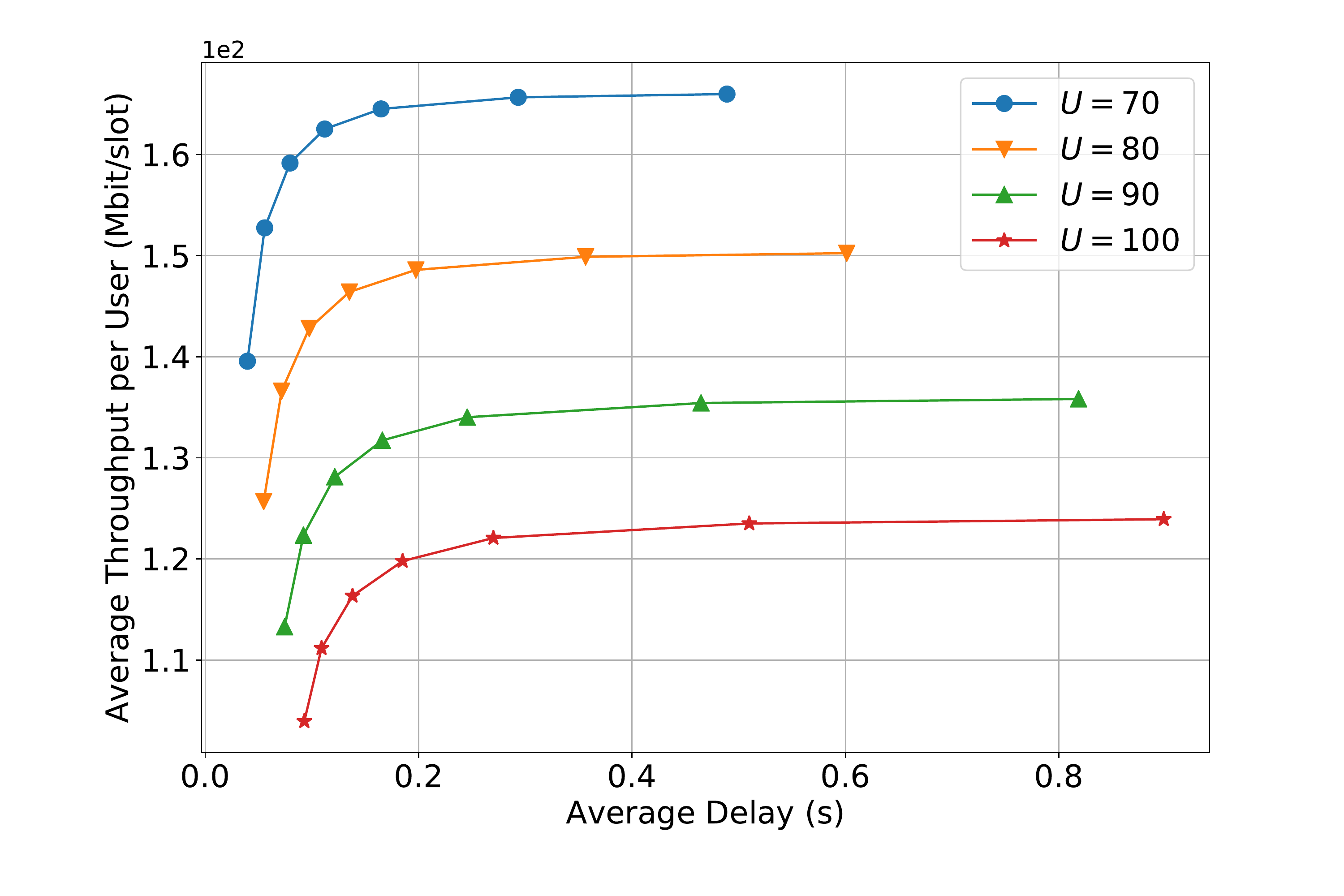}}\label{fig:AverageQueueDiffworkload}
   \end{minipage}
   \caption{
		Performance of PUARA under different maximum arrival rates and numbers of users. From the results, we see that: (a) a larger maximum arrival rate $A_{\text{max}}$ leads to a lower average throughput and a longer average delay; (b) the larger the number of users, the lower the average throughput and the longer the average delay of the system.}
   \label{fig:PUARA_DiffWorkload}
\end{figure}

\textbf{Performance under different system parameters:}
In Figures \ref{fig:PUARA_DiffWorkload} (a) and \ref{fig:PUARA_DiffWorkload} (b), we investigate the performance metrics (throughput and delay) incurred by PUARA under different settings of maximum arrival rate $A_{\text{max}}$ and number of users $U$, respectively.
Each data point in figures corresponds to the performance of PUARA under one particular value of $V$. 
The value of $V$ is set as $1, 200, 500, 1000, 2000, 5000, 10000$ from left to right, respectively.

Figure \ref{fig:PUARA_DiffWorkload} (a) shows that in general, as the value of $V$ increases, 
the curve of time-averaged total network throughput ascends and eventually converges while the average delay of the system keeps increasing. 
Such results show that PUARA achieves a better throughput performance but at the cost of a longer average delay of the system. In contrast, to achieve a shorter average delay, 
we can decrease the value of $V$ from $10^4$ to $2 \times 10^3$ so that 
only a mild amount of throughput needs to be traded off (\textit{e.g.}, $1.73\%$ decrease in the throughput for a $73.45\%$ reduction in the average delay when $A_{\max}=120$Mbits). 
Therefore, PUARA actually achieves a tunable throughput-delay tradeoff, which is consistent with our results in Theorem \ref{Theorem1}. 
Besides, we also see that the increase of the maximum arrival rate $A_{\text{max}}$ leads to a lower time-averaged network throughput. This is because as users request more files, they would cause more contention of service rates and hence a longer average delay of the system and a lower time-averaged throughput.
Likewise, Figure \ref{fig:PUARA_DiffWorkload} (b) shows an increased time-averaged throughput for each user and a decreased average delay for the system as the number of users increases.

\setlength{\textfloatsep}{2pt}

\begin{figure}[!t]
        \centering \includegraphics[width=0.75\columnwidth]{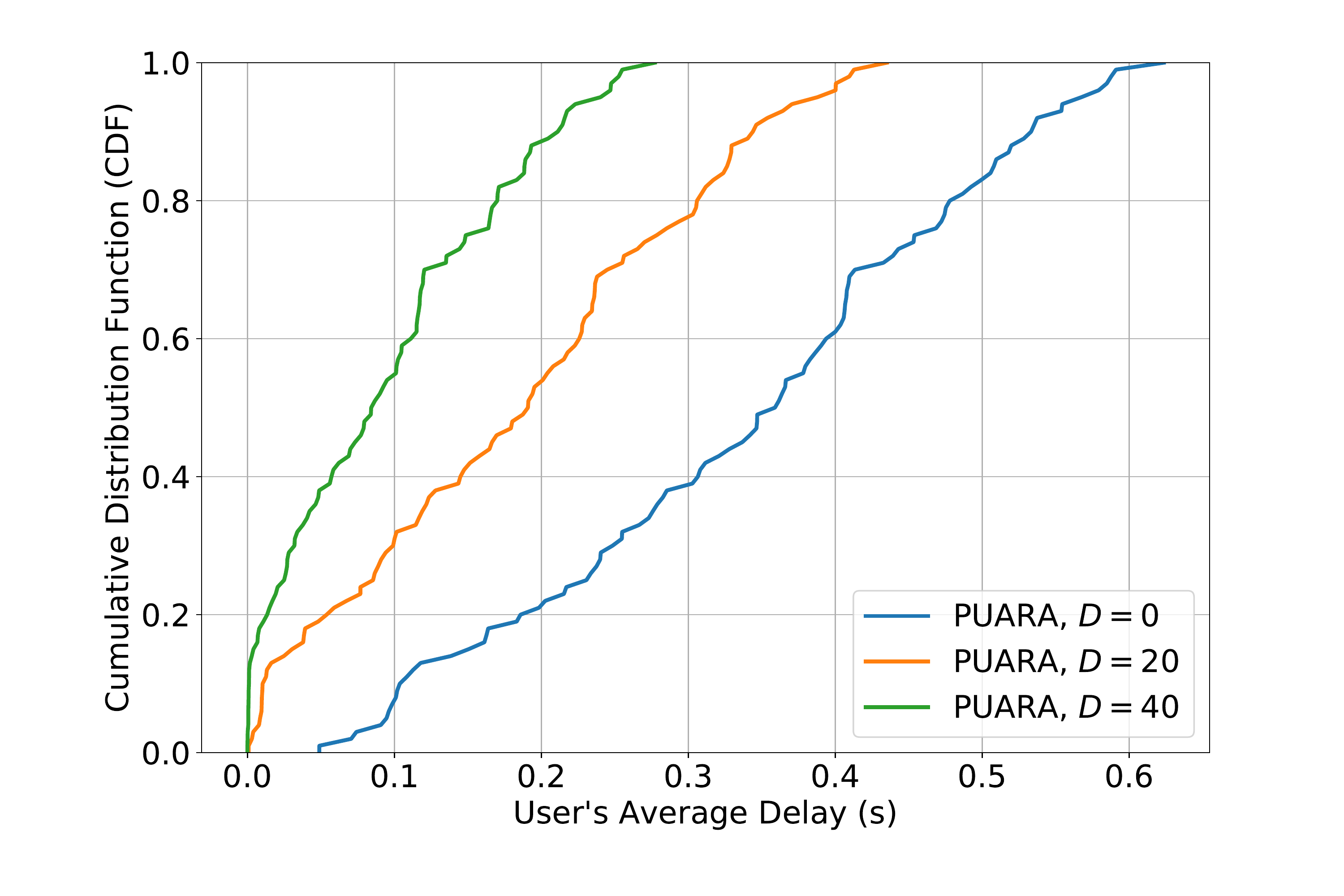}
        \caption{CDFs of each user's average delay under different values of window size $D$. 
        	Increasing window size $D$ incurs a notable left shift of the CDF curve (\textit{e.g.}, with a $31.65\%$ reduction for $95\%$ of users as the value of $D$ increases to $20$).} 
        \label{CDF}
\end{figure}

\textbf{Average delay of each user under different prediction window sizes:}
Figure \ref{CDF} shows the cumulative distribution functions (CDFs) of the average delay for each user\footnote{
		In our simulations, the average delay of each user $u$ is defined as 
		$\sum_{f\in\mathcal{F}} \bar{Q}_{uf} / \bar{A}_{u}$. By Little's theorem\cite{little1961proof}, this is equal to the average delay for user $u$ to transmit each Mbits of data in the system.} 
        under different values of prediction window size $D$ (with $V=1$). Note that each curve is drawn from the same population of $100$ users.
Particularly, we find a notable left shift of the CDF curve as the prediction window size $D$ increases (\textit{e.g.}, by an $31.65\%$ delay reduction for $95\%$ of users as the value of $D$ increases from $0$ to $20$).
Such results imply that only mild value of future information suffices to aid PUARA to incur a notable reduction in the average delay of each user.

\begin{figure}[!t]
        \centering
        \vspace{1em}
	\includegraphics[width=0.75\columnwidth]{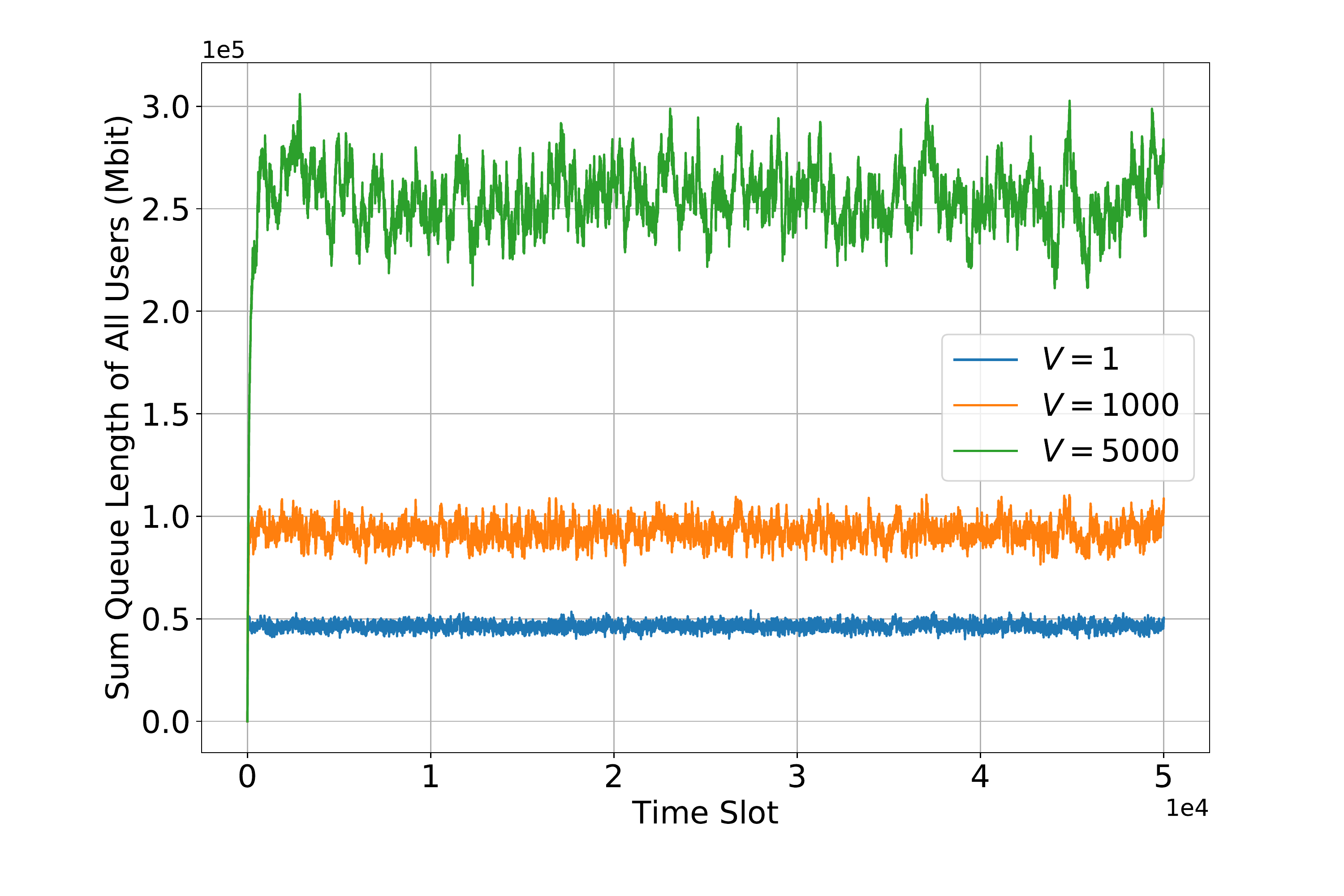}
        \caption{Total queue backlog sizes over time slots.
         All queue backlogs eventually stabilize with a total size proportional to the value of $V$.
    }\label{Q_t} 
\end{figure}

\textbf{Queue stability under different values of $V$:}
Figure \ref{Q_t} shows the variations of the total queue length in the system over time slots with $D=20$. 
Our results show that larger values of $V$ generally lead to relatively longer convergence times; moreover, 
PUARA can direct queueing dynamics in the system towards the stable state within hundreds of time slots (few seconds). Once entering the stable state, the total queue length remains fluctuating around a fixed level (\textit{e.g.}, $\pm 17.9\%$ around $2.57 \times 10^5$ when $V=5000$).
Instead of unbounded delays, by Little's law \cite{little1961proof}, such stability guarantee can ensure the timely processing of user requests.

\setlength{\textfloatsep}{3pt}

\begin{figure}[!t]
   \begin{minipage}[t]{1.0\linewidth}
        \centering
        \subfigure[Throughput/delay under file type mis-prediction]
        	{\includegraphics[width=0.7\columnwidth]{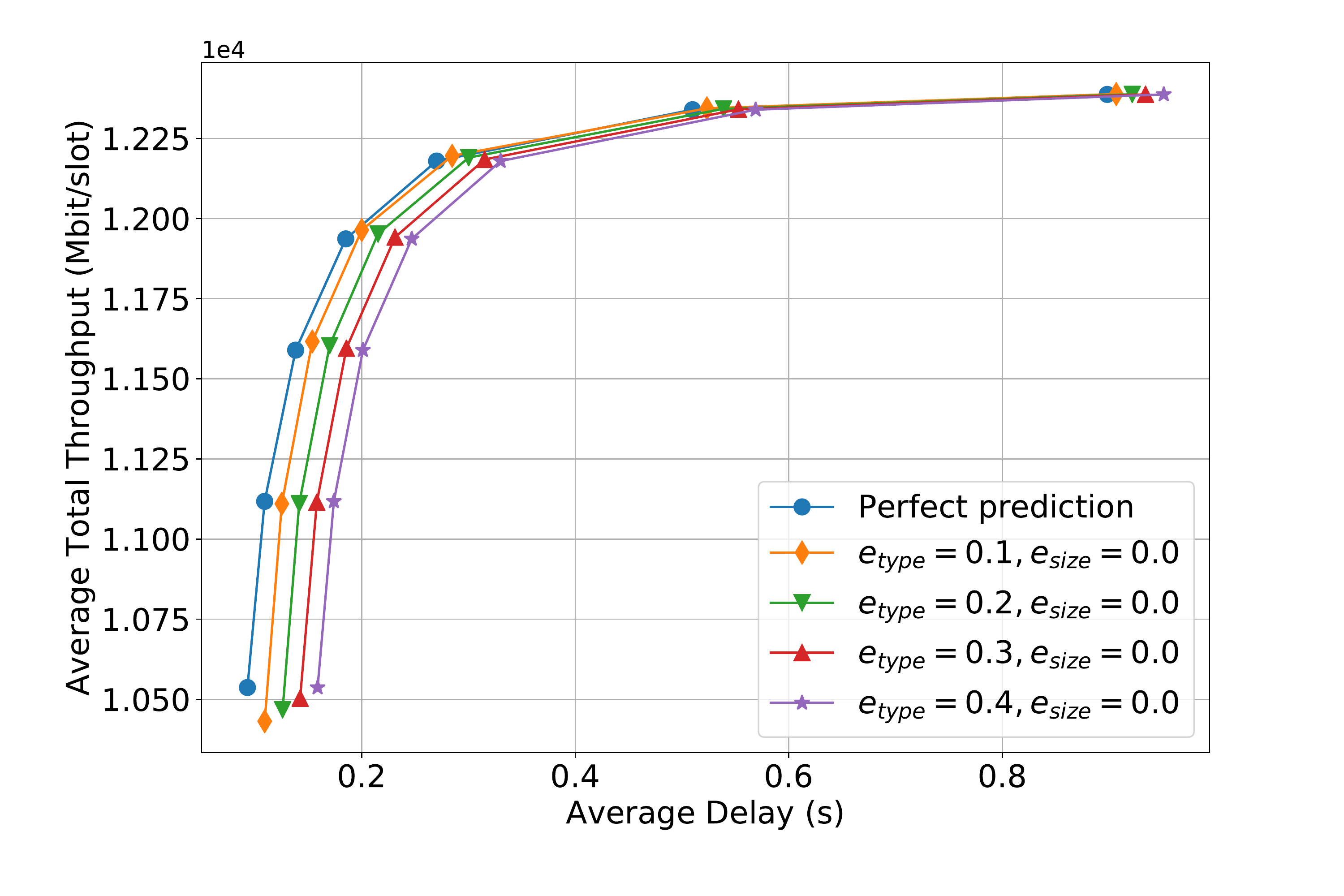}}
        \subfigure[Throughput/delay under file size mis-prediction]
        	{\includegraphics[width=0.7\columnwidth]{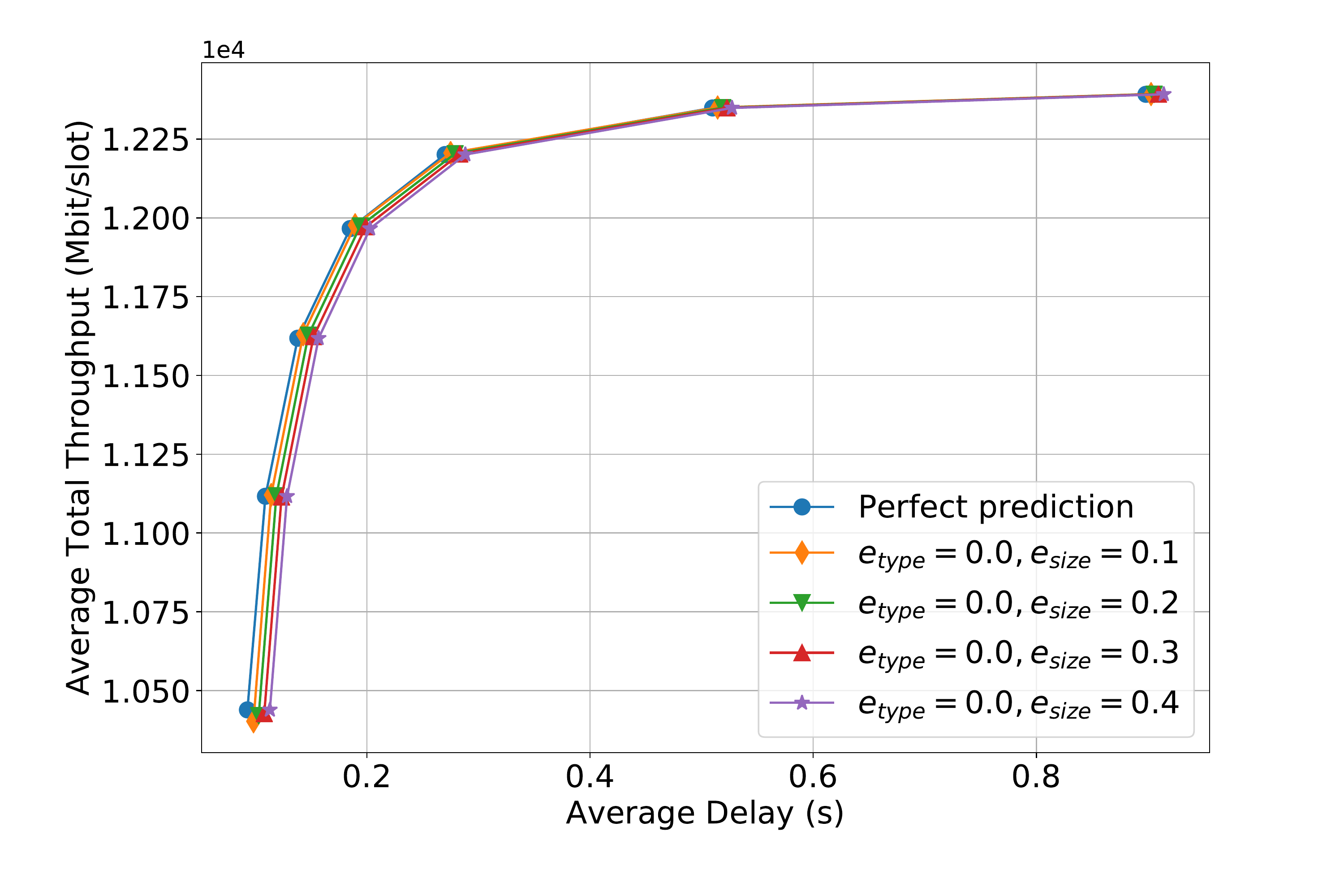}}
        \subfigure[Throughput/delay under two types of mis-prediction]{\includegraphics[width=0.7\columnwidth]{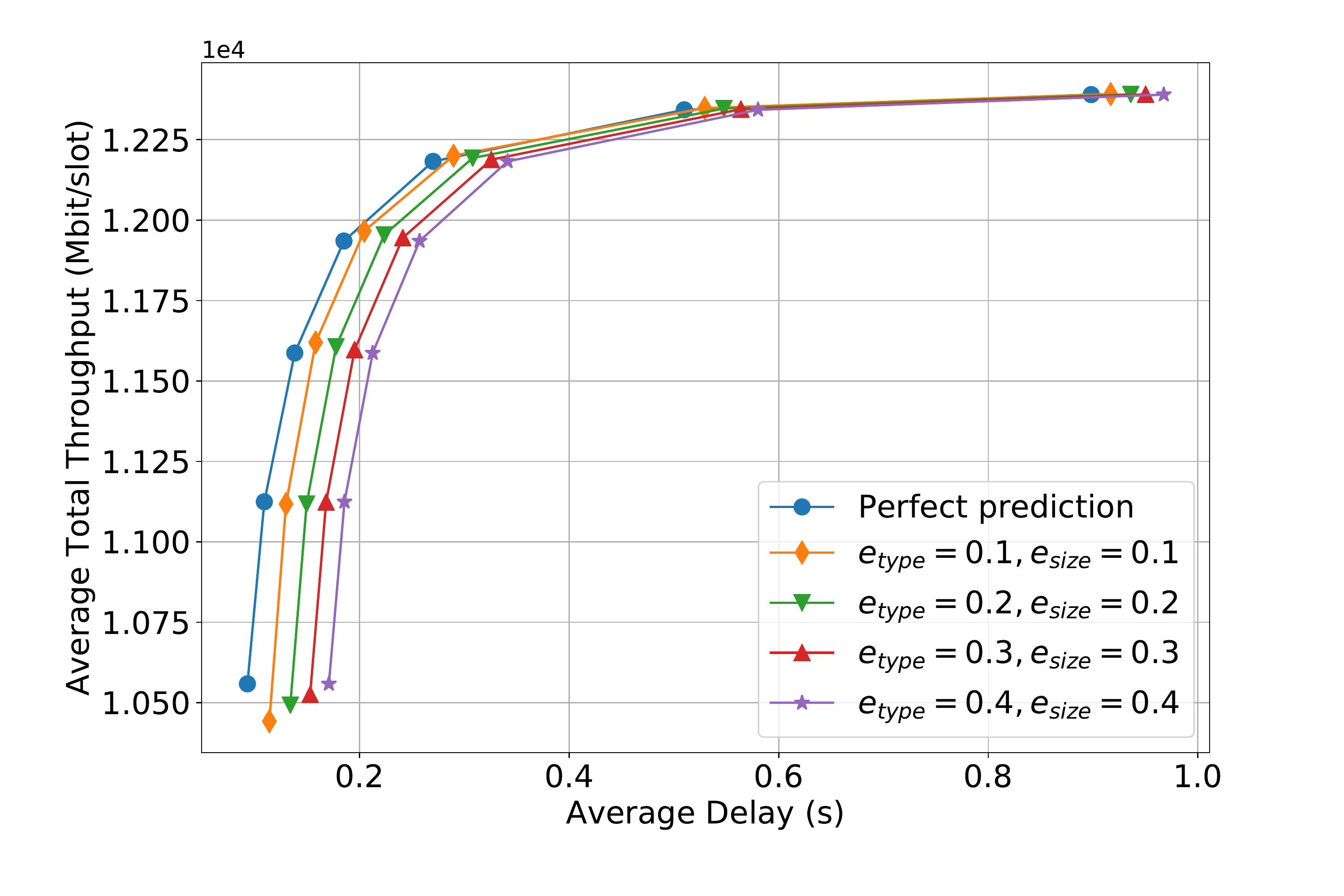}}
   \end{minipage}
   \caption{Performance of PUARA under different types of mis-predictions with $A_{\text{max}} = 100\text{Mbits}$ and $D=20$. 
   In general, we find that higher mis-prediction probabilities can lead to a lower throughput and a longer average delay of the system. However, as the value of $V$ increases, the impact of mis-prediction becomes gradually weakened. 
   Besides, compared to file size prediction, file type prediction has a greater impact on the time-averaged throughput and the average delay of the system.}
   \label{fig:PUARA_DiffPredictionErr}
\end{figure}
\setlength{\textfloatsep}{5pt}

\subsection{Performances under imperfect prediction}  \label{subsec: sim results under imperfect prediction}
In practice, mis-prediction are often inevitable.
In this part, we investigate the performance of PUARA under two cases of mis-prediction which are assumed independent of each other.
One is when a request's file type is mis-predicted. 
The other is when the total size of files requested by each user during some time slot is mis-predicted. 
Regarding file type mis-prediction, in our simulations, we assume that the file type of each request may be mis-predicted with a probability of $e_{\text{type}}$. Moreover, each mis-predicted request is equally likely to be wrongly categorized as one of the other $(F-1)$ types.
Regarding file size mis-prediction, for each requested file with size $s$, we assume that its size is mis-predicted as a value uniformly distributed over $[s \times (1-e_{\text{size}}), s \times (1+e_{\text{size}})]$ with $e_{\text{size}} \in [0, 1]$.
In our simulations, we use $e_{\text{type}}$ and $e_{\text{size}}$ to denote the prediction error rates for file type mis-prediction and file size mis-prediction, respectively.
Then for each mis-predicted request, it is handled as follows.

a) When its requested file's type is mis-predicted but not pre-served before its actual arrival, then it will be eliminated from its corresponding queue by the time slot it is predicted to arrive in. However, when the request is pre-served, then the downloaded files will be removed by its user if it is not used by the time it is predicted to arrive at.

b) When its requested file's size is mis-predicted but it is not pre-served before its actual arrival, then the request will be simply eliminated from its queue by the time slot it is predicted to arrive in. However, when the request is pre-served, then upon its actual arrival, the user will first check whether the file has been fully downloaded. If not, then extra service rates will be consumed to finish the transmission of such files first before serving other requests. 

Given the above description, we define the average throughput and the average delay of the system under mis-prediction scenarios as follows, respectively.

a) For average throughput, we define it as the time-averaged throughput incurred by the transmission of not only the actually requested files but also those predicted (including the mis-predicted) files. 

b) For average delay of the system, we adopt the foregoing definition $\sum_{u \in \mathcal{U}} \sum_{f\in\mathcal{F}} \bar{Q}_{uf} / (\sum_{u \in \mathcal{U}} \bar{A}_{u})$. Note that the amounts of mis-predicted files are also counted in the calculation of time-averaged queue length $\bar{Q}_{uf}$ and the average arrival rate $\bar{A}_{u}$.\footnote{
			Some mis-predicted requests may be pre-served (their requested files are pre-downloaded onto user devices). Such files may not be finally used and will stay on user devices until being replaced by other newly downloaded files. Accordingly, such files will consume extra service rates and prolong the delays of subsequent requests.} 

\textbf{Performance under fixed prediction error rates:} 
We present our simulation results (with window size $D=20$) in Figures \ref{fig:PUARA_DiffPredictionErr}.
Note that each point in the figures corresponds to the result under a given setting of $V$, whose values vary from $1$ to $10^4$.
In Figure \ref{fig:PUARA_DiffPredictionErr} (a), we examine the impact of file size mis-prediction by assuming the perfect prediction of requested file types. 
		Then in Figure \ref{fig:PUARA_DiffPredictionErr} (b), we show how file type mis-prediction affects the throughput and average delay of the system when the file size of each request is perfectly predicted. 
		In Figure \ref{fig:PUARA_DiffPredictionErr} (c), we investigate how both types of mis-prediction jointly affect system performances. 
We make the following observations.

	\textit{First}, higher prediction error rates generally lead to longer average delays of the system but with only mild change in the throughput. The reason is that in such cases, the system may also have to serve those mis-predicted requests, thereby lengthening the average delay. However, as PUARA inclines to greedily allocate service rates to the most loaded user queue in a dynamic fashion, thus mis-prediction only leads to mild change in the average throughput.
	\textit{Second}, compared to file size mis-prediction, file type mis-prediction has a greater impact on system performances. For example, in Figure \ref{fig:PUARA_DiffPredictionErr} (a), when $V=10^{4}$, as the value of prediction error rate $e_{\text{type}}$ increases from $0$ to $0.4$, the average delay of the system increases by $5.89\%$, while in Figure \ref{fig:PUARA_DiffPredictionErr} (b), the average delay (with $e_{type}=0$ and $e_{size}=0.4$) is only $1.78\%$ longer than the perfect prediction case. Intuitively, the reason is that if a request's file type is mis-predicted and pre-served, then its user needs to re-download its requested file upon its actual arrival. Compared to file size mis-prediction, such re-acquisition requires more service rates and thus causes longer delays to its subsequent requests. 
	\textit{Third}, as the value of $V$ increases from $1$ to $10^4$, the impacts of mis-prediction on the average throughput and the average delay of the system become gradually weakened. Specifically,
	\begin{itemize}
		\item[$\diamond$] under file type mis-prediction (with $e_{\text{type}}=0.4$), 
				the reduction percentages of the average throughput and the average delay decrease from $1.38\%$ to zero and from $70.76\%$ to $5.89\%$, respectively;
		\item[$\diamond$] under file size mis-prediction (with $e_{\text{size}}=0.4$),
				the reduction percentages of the average throughput and the average delay decrease from $0.44\%$ to zero and from $21.66\%$ to $1.78\%$, respectively.
	\end{itemize}
	The reason is that given a larger value of $V$, PUARA is more prone to connecting users to their nearby APs with the greatest service rates (so as to improve the average throughput), which conduces to maximizing the utilization of available service capacities and mitigating the impact of mis-prediction. In contrast, with a smaller value of $V$, PUARA puts more focus to balance the workloads among users. To this end, PUARA is more prone to first serving those most loaded users but probably with less effective utilization of system resources. 
	\textit{Fourth}, in Figure \ref{fig:PUARA_DiffPredictionErr} (c), when both types of mis-prediction are considered with $V=10^{4}$, as the values of $e_{\text{type}}$ and $e_{\text{size}}$ increase from $0$ to $0.4$, 
	PUARA achieves an increase in the average delay of the system by $7.71\%$ with insignificant variations in the average throughput. Such results demonstrate the robustness of PUARA in the scenarios with mis-prediction. 

\begin{figure}[!t]
        \centering
        \vspace{-0.6em}
	\includegraphics[width=0.78\columnwidth]{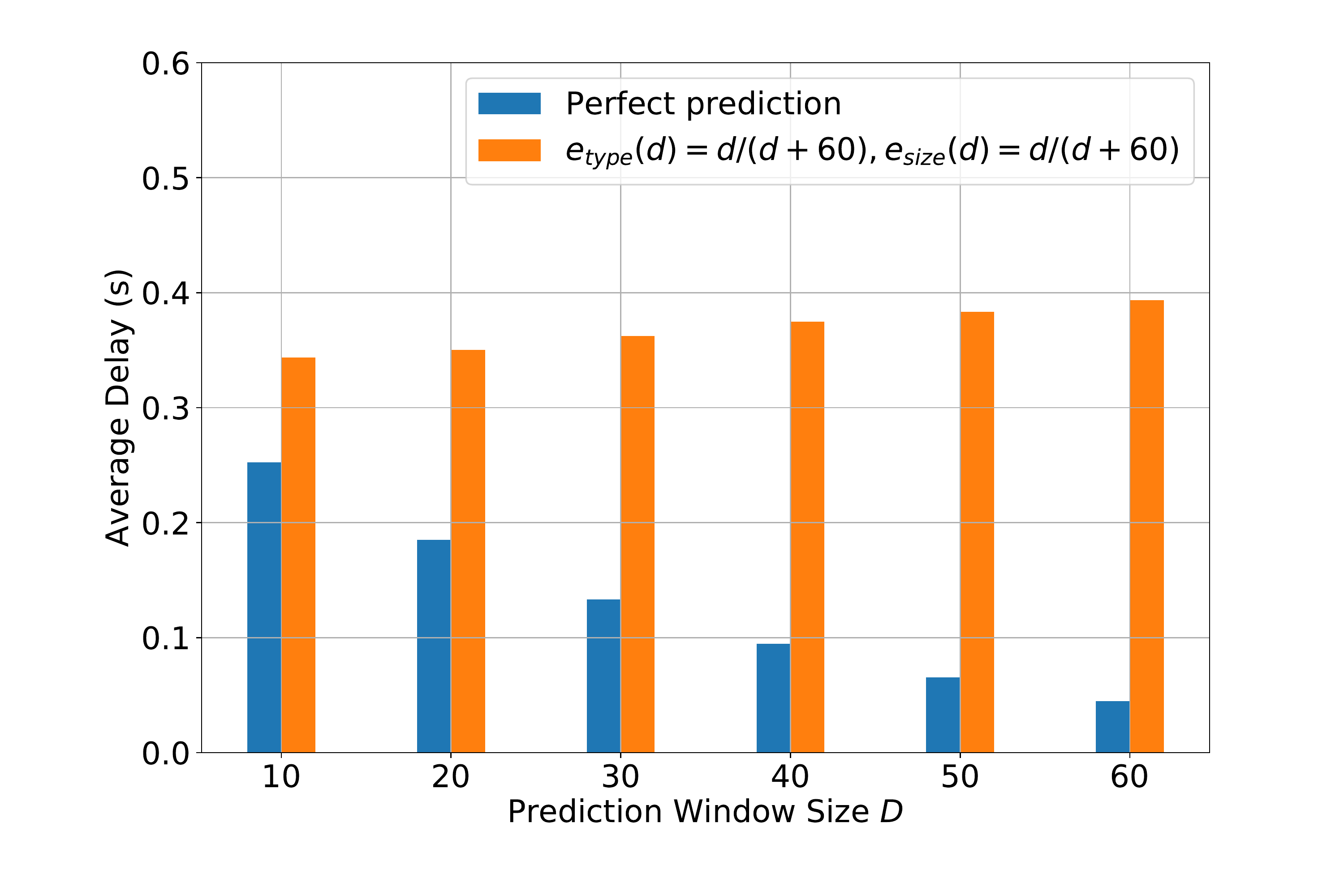}
	\vspace{-0.6em}
        \caption{Average delay of the system under time-varying prediction error rates with $V=1000$ and different values of window size $D$.
    }\label{fig:PUARA_DiffPredictionErrTV} 
\end{figure}

\textbf{Performance under time-varying prediction error rates:} 
In Figure \ref{fig:PUARA_DiffPredictionErrTV}, we investigate system performances under different time-varying mis-prediction settings.
We assume that prediction error rates $e_{\text{type}}(d)$ and $e_{\text{size}}(d)$ increase with the number of slots away from current time slot. 
Specifically, in our simulations, for the $d$-th time slot in the prediction window, we set $e_{\text{type}}(d)=\frac{d}{d+C}$ and $e_{\text{size}}(d)=\frac{d}{d+C}$ (both with $C=60$), respectively.  
Intuitively, under such settings, the prediction error rates for upcoming slots are lower than that of those farther slots. 
Note that for $Q^{d}_{uf}(t)$ ($0 \leq d \leq D-1$), if the predicted file type in time slot $t$ is updated from type $f$ to type $f'$ by time $(t+1)$, then $Q^{d}_{uf}(t+1)$ and $Q^{d}_{uf'}(t+1)$ will be updated as zero and the predicted size, respectively. 
Besides, for each time slot $(t+d)$ ($0 \leq d \leq D-1$) in the time window, when the predicted total size of requested files is updated in time slot $t$, 
the size of files that are already downloaded in previous time slots will be deducted from the predicted size. 

We first see that compared to the perfect prediction case, the increase in the average delay of the system due to mis-prediction ascends from $0.09$s to (at most) $0.35$s as the window size $D$ increases from $10$ to $60$. The reason is that the system has to allocate extra service rates to mis-predicted requests, which prolongs the delays of actual requests. 
Second, under mis-prediction, the increase in the average delay of the system is about $14.5\%$ as the prediction window size $D$ increases from $10$ to $60$. 
By comparing such results with Figure \ref{fig:PUARA_DiffPredictionErr} (c) in which error rates are fixed across time slots with $D=20$ and $V=1000$, we see that the average delay of the system is longer when prediction error rates $e_{\text{type}}$ and $e_{\text{size}}$ vary across time slots. 
The reason is that when error rates vary over time slots, within the prediction window of each user, there will more mis-predicted requests in the remote time slots than the upcoming slots. 
As a result, compared to the case with fixed error rates, the system may wrongly allocate service rates to more mis-predicted requests. Even though the number of mis-predicted requests in each time slot generally keeps decreasing as the prediction window advances, the consequence of such mis-allocation (in terms of long delays for subsequent requests) cannot be restored. 
Such results demonstrate the limits of the benefits of predictive scheduling. 
In practice, to exploit such benefits, system designers should choose a proper window size $D$ such that prediction error rates do not vary significantly across time slots in the time window.

\setlength{\textfloatsep}{3pt}

\section{Conclusions}\label{Conclusions}
In this paper, we studied the problem of joint user-AP association and resource allocation for content delivery with predictive scheduling over a fixed content placement. 
We devised an effective predictive scheme which achieves a provably near-optimal throughput with queue stability. 
Then we investigated the fundamental limits of benefits from predictive scheduling through theoretical analysis and simulations. Our results show that our scheme not only performed a tunable control between throughput maximization and queue stabilization but also incurred a notable delay reduction when given predicted information. 
In addition, some interesting directions can be left for future work, \textit{e.g.}, how to take into account the fairness of resource allocation among users and how to leverage user mobility dynamics to further improve system performance.






\appendices

\section{}
\label{appdix2}
According to constraint (\ref{ChannelRateConstraint2}), we know that only when user $u$ associates with the AP $h$ ($X_{uh} = 1$), 
the user will be allocated bandwidth by the AP $h$ \textit{i.e.}, $\nu_{uh} \geq 0$ and $0$ otherwise. Accordingly, problem $\mathcal {P}_{in}$ can be reduced to

\begin{eqnarray}\label{sub_objectiion2}
     \max\limits_{\boldsymbol{\nu}(t)} && \sum_{h \in \mathcal {H}}\sum_{u \in \mathcal {U}}\mathcal {M}_{uh}(t)\nu_{uh}(t) \label{ReducedProblem} \\
   \nonumber \text{s.t.} && (\ref{constraint: number of users for each AP}),(\ref{constraint: number of associated APs}), \\
   && \boldsymbol{X}(t) \in \{0, 1\}^{U \times H}, \\
   && \sum_{u \in \mathcal {U}}\nu_{uh} = 1 \  \forall h \in \mathcal{H},\\
   && \nu_{uh} \geq 0, \text{ if } X_{uh} = 1 \ \forall u \in \mathcal{U}, h \in \mathcal{H}, \\
   && \nu_{uh} = 0, \text{ if } X_{uh} = 0 \  \forall u \in \mathcal{U}, h \in \mathcal{H}.
\end{eqnarray}

Next, through proof by contradiction, we prove that each AP $h \in \mathcal {H}$ associates with only one user in time slot $t$ for the optimal user-AP association $\boldsymbol {X}^\star(t)$.

We first assume that under the optimal user-AP association $\boldsymbol {X}^\star(t)$, the AP $h$ associates with more than one user, \textit{i.e.}, $X_{u_1 h}^\star = 1$, $X_{u_2 h}^\star = 1$, $X_{uh}^\star = 0$ where $u \in \mathcal {U}\backslash \{u_1,u_2\}$. Thus we have $\nu_{u_1 h}^\star(t) \geq 0$, $\nu_{u_2 h}^\star(t) \geq 0$, $\nu_{u_1 h}^\star(t) + \nu_{u_2 h}^\star(t) = 1$ and $\nu_{uh}^\star(t) = 0$ where $u \in \mathcal {U}\backslash \{u_1,u_2\}$.

We also assume that $\mathcal {M}_{u_1h}(t) \geq \mathcal {M}_{u_2h}(t)$.
Let $x'_{u_1 h}(t) = 1$, $x'_{u_2 h}(t) = 0$ and $\nu'_{u_1 h}(t) = \nu_{u_1 h}^\star(t) + \nu_{u_2 h}^\star(t) = 1$, $\nu'_{u_2 h}(t) = 0$, then we have that:
\begin{eqnarray}
  \nonumber && \mathcal {M}_{u_1h}(t) = \mathcal {M}_{u_1h}(t)\nu'_{u_1h}(t) \\
   \nonumber &\geq& \mathcal {M}_{u_1h}(t)\nu_{u_1 h}^\star(t) + \mathcal {M}_{u_2h}(t)\nu_{u_2 h}^\star(t),
\end{eqnarray}
Therefore, $x'_{u_1 h}(t) = 1$, $x'_{u_2 h}(t) = 0$ and $\nu'_{u_1 h}(t) = 1, \nu'_{u_2 h}(t) = 0$ are the optimal solution of (\ref{ReducedProblem}), where the AP $h$ associates with the user $u$ with the maximum $\mathcal {M}_{uh}(t)$, and allocates the whole bandwidth to it, which is contradictory with the assumption before.
Extending this to all APs, then the optimal user-AP association can be obtained by solving the following problem:
\begin{eqnarray}\label{sub_objectiion3}
     \max\limits_{\boldsymbol{X}(t)} && \sum_{h \in \mathcal {H}}\sum_{u \in \mathcal {U}}\mathcal {M}_{uh}(t)X_{uh}(t)\\
   \nonumber \text{s.t.} && \sum_{u \in \mathcal {U}} X_{uh}(t) \leq M \ \forall h \in \mathcal {H}, \\
   && \sum_{h \in \mathcal {H}} X_{uh}(t) \leq 1\ \forall u \in \mathcal {U},\\
   && \boldsymbol{X}(t) \in \{0,1\}^{U \times H},
\end{eqnarray}
and the corresponding optimal solution $\boldsymbol{\nu}(t) = \boldsymbol{X}(t)$. Hence, problem $\mathcal {P}'_{in}$ is equivalent to problem $\mathcal {P}_{in}$.
\IEEEQED

\section{}
\label{matroid-transformation}
To conduct the problem transformation, we introduce a ground set, denoted by $\mathcal{X}$, which consists of all possible pairs of user-AP association, \textit{i.e.},
\begin{eqnarray*}
   \mathcal{X} \triangleq \{\mathcal {X}_1^1,\mathcal {X}_2^1,\ldots,\mathcal {X}_U^1,\ldots,\mathcal {X}_1^H,\mathcal {X}_2^H,\ldots,\mathcal {X}_U^H\},
\end{eqnarray*}
where $\mathcal {X}_u^h$ denotes the association between user $u$ and AP $h$. For each set $\underline{\mathcal {X}} \subseteq \mathcal {X}$, we use $|\underline{\mathcal {X}}|$ to denote its cardinality. 

Regarding the constraints in problem $\mathcal {P}'_{in}$, the following lemma reveals its unique structure in terms of matroids.
\begin{lemma}
\textit{
The constraints of problem $\mathcal {P}'_{in}$ can be written as the intersection of two partition matroids $\mathcal {A}_1 = (\mathcal {X},\mathcal {I}_1)$ and $\mathcal {A}_2 = (\mathcal {X},\mathcal {I}_2)$ over ground set $\mathcal {X}$.	
}
\end{lemma}
The proof is relegated to Appendix \ref{appdix3}.

Meanwhile, We can rewrite the objective function of problem $\mathcal {P}'_{in}$ as
\begin{equation}\label{ModularFunction}
  g(\underline{\mathcal {X}}) = 
  \sum_{u \in \mathcal {U}}
  \sum_{h \in \mathcal {H}: \mathcal {X}_u^h \in \underline{\mathcal {X}}} 
  	\mathcal{M}_{uh}(t)\mathcal {X}_u^h.
\end{equation}
The following lemma shows that $g(\cdot)$ is a modular function.
\begin{lemma}
	\textit{
	Let $\mathcal {X}_1 \subset \mathcal {X}_2 \subset \mathcal {X}$ and $\mathcal {X}_u^h \in \mathcal {X} - \mathcal {X}_2$, then $g(\cdot)$ is a modular function, \textit{i.e.}, 
\begin{equation}\label{ModularLemma}
  g(\mathcal {X}_1 \cup \mathcal{X}_u^h) - g(\mathcal {X}_1) = g(\mathcal {X}_2 \cup \mathcal {X}_u^h) - g(\mathcal {X}_2).
\end{equation}}
\end{lemma} 
The proof is relegated to Appendix \ref{appdix4}.

Therefore, problem $\mathcal {P}'_{in}$ is equivalent to a modular maximization problem over two intersected matroid constraints.
\IEEEQED

\section{}
\label{appdix3}
The ground set $\mathcal {X}$ can be partitioned into $H$ disjoint subsets: $S_1,\ldots,S_H$, where $S_h = \{\mathcal {X}_1^h,\ldots,\mathcal {X}_U^h\}$ is the set of all users that might associate with AP $h$.
Remind that the user-AP association is expressed by the matrix $\boldsymbol {X}(t)$. We define the user-AP association set $X \subseteq \mathcal {X}$ such that $\mathcal {X}_u^h \in X$ if and only if $X_{uh}(t)  =1$. Notice that the nonzero elements of the $h$th column of matrix $\boldsymbol{X}(t)$ equals to the elements in $X \subseteq S_h$. Thus the constraint of the column of matrix $\boldsymbol {X}(t)$ can be expressed as $X \subseteq \mathcal {I}_1$, where
\begin{eqnarray}
   \mathcal {I}_1 = \{X \subseteq \mathcal {X}: |X \cap S_h| \leq M, \forall h = 1,\ldots,H\}.
\end{eqnarray}
Comparing $\mathcal {I}_1$ with the definition of partition matroid\cite{edmonds1971matroids}, we see that constraints in (\ref{constraint: number of users for each AP}) form a partition matroid with $l = H$ and $k_i = 1$ for $i = 1,\ldots,H$. We denote this partition matroid by $\mathcal {A}_1 = (\mathcal {X},\mathcal {I}_1)$.

Similarly, the ground set $\mathcal {X}$ can also be partitioned into $U$ disjoint subsets: $S'_1,\ldots,S'_U$, where $S'_u = \{\mathcal {X}_u^1,\ldots,\mathcal {X}_u^H\}$ is the set of all AP that might associate with user $u$. The constraint of the row of matrix $\boldsymbol {X}(t)$ can be expressed as $X \subseteq \mathcal {I}_2$, where
\begin{eqnarray}
   \mathcal {I}_2 = \{X \subseteq \mathcal {X}: |X \cap S'_u| \leq 1, \forall u = 1,\ldots,U\}.
\end{eqnarray}

Hence, constraints in (\ref{constraint: number of associated APs}) form a partition matroid with $l = U$ and $k_i = 1$ for $i = 1,\ldots,U$. This partition matroid is denoted by $\mathcal {A}_2 = (\mathcal {X},\mathcal {I}_2)$.

To sum up, the constraint of problem $\mathcal {P}'_{in}$ can be expressed as two partition matroids on a ground set $\mathcal {X}$.
\IEEEQED

\section{}
\label{appdix4}
It's easily to be verified that the function $h(x) = x$ is strictly increasing and linear for all $x > 0$. This means that we have
\begin{equation}\label{LinearFunction}
    h(x+1) - h(x) = h(y+1) -h(y), \ \forall 0 < x < y
\end{equation}
Combining (\ref{LinearFunction}) with the facts that $h(0) = 0$ and $f(\underline{\mathcal {E}}) = h(|\underline{\mathcal {E}}|)$ (so that $f({\o}) = 0$, where ${\o}$ is empty set), we have
\begin{eqnarray}
  \nonumber  f(\underline{\mathcal {E}} \cup (u,h)) - f(\underline{\mathcal {E}}) = f(\underline{\mathcal {E}}^{'}
  \cup (u,b)) - f(\underline{\mathcal {E}}^{'}), \\
   \forall  \underline{\mathcal {E}} \subseteq \underline{\mathcal {E}}^{'} \subseteq \mathcal {E} \& (u,b)
   \in \mathcal {E} \backslash \underline{\mathcal {E}}^{'} \label{SubmodularProof}
\end{eqnarray}
which yields the desired result. Hence, the objective function in Problem $\mathcal {P}_{in}$ is a modular function.
\IEEEQED

\section{}
\label{imperfect scheduling}
We define the following problem as a good reference point for the solution under imperfect scheduling.

$\boldsymbol{\beta}$-\textbf{reduced problem}:
\begin{eqnarray}
   \text{max} && \phi(\overline{\boldsymbol{\mu}}) \label{BetaReducedProb} \\
  \text{s.t.} && \boldsymbol{\mu}(t) \in \beta\mathcal {R}, \label{NewRateConstraint}\\
  &&\overline{Q}_{uf} < \infty, \ \forall u \in \mathcal {U}, f \in \mathcal{F}, \\
  &&\alpha(t) \in \mathcal {A}_{\omega(t)}, \ \forall\,t,
\end{eqnarray}
in which we recall that $\phi$ is a linear function.

Next, we define $\boldsymbol{\mu}^{*,\beta}(t)$ as the optimal solution to $\beta$-reduced problem and $\boldsymbol{\mu}^{*,0}(t)$ as the optimal solution to problems $\mathcal{P}_{2}$ and $\mathcal {P}_{in}$.
The following lemma establishes the relationship between problem $\mathcal{P}1$ and $\beta$-reduced problem in terms of their optimal solutions.
\begin{lemma}\label{beta-reduced}
\textit{
 Let $\boldsymbol{\mu}^{*,0}(t)$ be the optimal solution of the $\mathcal {P}1$. Then the solution to the $\beta$-reduced problem is
\begin{equation}\label{WeightedProportionFair}
    \boldsymbol{\mu}^{*,\beta}(t) = \beta\boldsymbol{\mu}^{*,0}(t).
\end{equation}}
\end{lemma}
The proof is relegated to Appendix \ref{appdix5}. 

\section{}
\label{appdix5}
In $\beta$-reduced problem (\ref{BetaReducedProb}), by a change of variables $\boldsymbol{\mu}'(t) = \boldsymbol{\mu}(t)/\beta$ and the fact that
$    \phi(\boldsymbol{\mu}(t)) = \sum_{u \in \mathcal {U}}\mu_u$,
we have
\begin{equation}
    \phi(\mu_u(t)) = 1/\beta \mu'_u(t).
\end{equation}
Then it follows that $\beta$-reduced problem becomes equivalent to the problem $\mathcal {P}1$. Hence,
$\boldsymbol{\mu}^{*,\beta}(t) = \beta\boldsymbol{\mu}^{*,0}(t) $.
\IEEEQED

\section{}\label{appdix8}
To characterize the performance of PUARA, we assume that all the random events in the system are \textit{i.i.d.} and the following slater-type conditions hold.
\begin{eqnarray}
   && \lambda_u - \sum_{\omega_j}\pi_{\omega_j}\sum_{m}\varphi_m^{\omega_j}\mu_u(\alpha_m^{\omega_j}) \leq -\theta,   \  \forall u \in \mathcal {U} \label{PredictiveSlaterCondition1}\\
   && \phi(\boldsymbol{\mu}(\alpha_m^{\omega_j})) = \phi_{\theta},\label{PredictiveSlaterCondition2}
\end{eqnarray}
where $\theta \in (0,\epsilon]$ and $\phi_{\theta}$ is a finite constant. 
Note that $\theta \rightarrow 0$ as $D_u \rightarrow \infty$.
Then we have the following result
\begin{equation}\label{ThetaTo0}
  \lim_{\theta \rightarrow 0}\phi_{\theta} = \phi^{\text{opt}}.
\end{equation}
The above assumptions ensure strong stability of the queue backlogs in the system and the existence of at least one stationary and randomized policy.

To proceed, we introduce the following lemma which can be easily proved by applying (\ref{PredictiveMinimizedrift-plus-penaltyProblem}) under the \emph{slater-type condition} (\ref{PredictiveSlaterCondition1}) and (\ref{PredictiveSlaterCondition2}).
\begin{lemma}
\textit{
For any alternative policy $\alpha^{\omega_j} \in \mathcal {A}_{\omega}$, we have
\begin{equation}\label{Lemma5PredictiveNewDriftBound}
  \Delta_V^p(\boldsymbol{Q}(t)) \leq \mathcal {K} - V\phi_{\theta} - 
  \beta  \theta\sum_{u \in \mathcal {U}}\sum_{f \in \mathcal {F}}\sum_{d = -1}^{D_u -1}\tilde{Q}_{uf}^d(t).
\end{equation}}
\end{lemma}
Next, we define a quadratic Lyapunov function as
\begin{eqnarray}
  \nonumber   && L(\boldsymbol{Q}(t+1)) - L(\boldsymbol{Q}(t)) \\
  \nonumber & = & \frac{1}{2} \bigg\{ \big([\boldsymbol{Q}(t) - \boldsymbol{\mu}(t)]^{+}+ \boldsymbol{A}(t) \big)^{\text{T}}\big([\boldsymbol{Q}(t) - \boldsymbol{\mu}(t)]^{+} \\
&&  + \boldsymbol{A}(t) \big)- \boldsymbol{Q}^{\text{T}}(t)\boldsymbol{Q}(t) \bigg\}.  \label{LyapvDrift}
\end{eqnarray}
Next, we define the one-time-slot conditional Lyapunov \emph{drift-plus-penalty} function as
\begin{eqnarray}\label{ConditionLyapnvDrftPenlty}
    \Delta_V(\boldsymbol{Q}(t)) \triangleq 
    \Delta(\boldsymbol{Q}(t))
     - V\mathbb{E}\{\phi(\boldsymbol{\mu}(t))|\boldsymbol{Q}(t)\},
\end{eqnarray}
where $V$ is a positive parameter and $\Delta(\boldsymbol{Q}(t))$ is defined as
\begin{equation}\label{ConditionLyapnvDrft}
    \Delta(\boldsymbol{Q}(t))\triangleq \mathbb{E}\{L(\boldsymbol{Q}(t+1)) - L(\boldsymbol{Q}(t))|\boldsymbol{Q}(t) \}.
\end{equation}
Then we have the following lemma (see proof in Appendix \ref{appdix6}).
\begin{lemma}\label{l6}
For any feasible decision $\alpha(t)$ for $\mathcal {P}1$, $\Delta_V(\boldsymbol{Q}(t))$ is upper bounded (with $\mathcal {K} = \frac{U}{2}[\mu_{\text{max}}^2 + A_{\text{max}}^2]$) by
\begin{eqnarray}
    \nonumber \Delta_V(\boldsymbol{Q}(t))  & \leq & \mathcal{K}-V\mathbb{E}\{\phi(\boldsymbol{\mu}(t))|\boldsymbol{Q}(t)\} \\
     &+& \mathbb{E}\{\big(
     \boldsymbol{A}(t)-\boldsymbol{\mu}(t)\big)^{\text{T}}\boldsymbol{Q}(t)|\boldsymbol{Q}(t) \}.
     \nonumber
     \label{DrftPentlyBound}
\end{eqnarray}
\end{lemma}

\emph{Lemma} \ref{l6} provides an upper bound for the conditional Lyapunov \emph{drift-plus-penalty} function $\Delta_V(\boldsymbol{Q})$, which plays a significant role in analyzing PUARA. 
Our control policy 
aims to make decision $\boldsymbol{\alpha}(t) \in \boldsymbol{\mathcal {A}}_{\omega}(t)$ to minimize the upper bound of $\Delta_V(\boldsymbol{Q}(t))$, as shown in \eqref{drift-plus-penalty}.

To proceed, we apply the following lemma whose proof is relegated to Appendix \ref{appdix7}.
\begin{lemma}\label{l7}
For any alternative policy $\alpha^{\omega_j} \in \mathcal {A}_{\omega}$, we have:
\begin{equation}\label{Lemma4NewDriftBound}
  \Delta_V(\boldsymbol{Q}(t)) \leq \mathcal {K} - V\phi_{\epsilon} - 
  \beta\epsilon \sum_{u \in \mathcal {U}}\sum_{f \in \mathcal {F}}Q_{uf}(t).
\end{equation}
\end{lemma}
Next, by taking expectation of (\ref{Lemma4NewDriftBound}),
summing over $\tau \in \{0,1,\cdots,t-1\}$ for some slot $t > 0$,
then dividing the result by $t\epsilon$, and taking the limsup of both sides,
we obtain
\begin{eqnarray}
   \limsup_{t \rightarrow \infty} \frac{1}{t} \sum_{\tau = 0}^{t-1}\sum_{u}\sum_{f}\mathbb{E}\{Q_{uf}(t)\}
  \leq  \frac{\mathcal {K}+V(\phi^{\text{max}} - \phi_{\epsilon})}{
  \beta\epsilon}.
\end{eqnarray}
Next, we consider policy $\alpha^{\omega_j}(t)$ which achieves the optimal value $\phi_{\beta}^{\text{opt}}$ of the $\beta-$reduced problem $\mathcal {P}1$. We have
\begin{flalign}
    &\nonumber  \mathbb{E}\{L(\boldsymbol{Q}(\tau+1)) - L(\boldsymbol{Q}(\tau))\} -  V\mathbb{E}\{\phi(\boldsymbol{\mu}(\tau))\} \leq 
    \mathcal {K} - V \beta \phi^{\text{opt}}.
\end{flalign}
where $\beta = 1/2$, $\phi^{\text{opt}}$ is the optimal throughput.
Next, by summing the above over $\tau \in \{0,1,\cdots,t-1\}$, 
and dividing by $tV$ and rearranging terms, we have
\begin{equation}
	\begin{array}{cl}\label{MiddleIneq}
		& \displaystyle
		\frac{1}{t}\sum_{\tau=0}^{t-1}\mathbb{E}\{\phi(\boldsymbol{\mu}(\tau))\}
		\geq  
		\beta \phi^{\text{opt}}- \frac{\mathcal {K}}{V},
	\end{array}
\end{equation}
where the second inequality is due to the non-negativeness of $\mathbb{E}\{L(\boldsymbol{Q}(t))\}$ and $\mathbb{E}\{L(\boldsymbol{Q}(0))\}$.
Finally, by taking the lim-inf as $t \to \infty$, we have
\begin{equation}
    \liminf_{t \to \infty} \frac{1}{t} \sum_{\tau = 0}^{t-1}\mathbb{E}\{\phi(\boldsymbol{\mu}(\tau))\}
    \geq 
    \beta \phi^{\text{opt}} - \frac{\mathcal{K}}{V}.
\end{equation}
\IEEEQED

\section{}
\label{appdix6}
Note that for any $Q \geq 0, \mu \geq 0, A \geq 0$, we have
\begin{equation}\label{MidIneq1}
    \big([Q-\mu]^{+}+A \big)^2 \leq Q^2 + A^2 + \mu^2 +2Q(A-\mu).
\end{equation}
By applying (\ref{MidIneq1}) to the Lyapunov drift function (\ref{LyapvDrift}), we have
\begin{equation}
	\begin{array}{cl} \label{BoundDeprive}
		& L(\boldsymbol{Q}(t+1)) - L(\boldsymbol{Q}(t)) \\
		\leq & \frac{1}{2} 
		  \boldsymbol{A}(t)^{\text{T}}\boldsymbol{A}(t) + \big( 
		  \boldsymbol{A}(t)-\boldsymbol{\mu}(t)\big)^{\text{T}}\boldsymbol{Q}(t)  
		  + \frac{1}{2}\boldsymbol{\mu}(t)^{\text{T}}\boldsymbol{\mu}(t)\\
		\leq & \mathcal {K} + \big(  
		\boldsymbol{A}(t)-\boldsymbol{\mu}(t)\big)^{\text{T}}\boldsymbol{Q}(t), 	
	\end{array}	
\end{equation}
where $\mathcal {K} = \frac{U}{2}[\mu_{\text{max}}^2 + A_{\text{max}}^2]$ and $\mathbf{A}(t)$ refers to the queue backlog vector with future arrivals. 

By taking conditional expectation of (\ref{BoundDeprive}) and adding the penalty term $-V\mathbb{E}\{\phi(\boldsymbol{\mu}(t))|\boldsymbol{Q}(t)\}$ to both sides, we complete the proof of \emph{Lemma} \ref{l6}. 
\IEEEQED
\section{}\label{appdix7}
Recall the key idea of Lyapunov optimization of minimizing the right-hand-side of (\ref{DrftPentlyBound}).
Thus for any alternative (possibly randomized) imperfect policy $\alpha^{\omega_j} \in \mathcal {A}_{\omega}$, we have
\begin{flalign}
 \nonumber & \Delta_V(\boldsymbol{Q}(t)) \leq  \mathcal {K} - V\phi(\boldsymbol{\mu}^{*,\beta}) \\
 & +\sum_{u \in \mathcal {U}}\sum_{f \in \mathcal {F}}Q_{uf}(t)\mathbb{E}\{
 A_u(t)I_{uf}(t) - \mu_u^{*,\beta}(t)|\boldsymbol{Q}(t)\} \label{DPPTheorem1}
\end{flalign}
where $\boldsymbol{\mu}^{*,\beta} = (\mu_1^{*,\beta},\cdots,\mu_U^{*,\beta})$ are given by the imperfect scheduling policy $\alpha^{\omega_j} \in \mathcal {A}_{\omega}$.

By applying (\ref{WeightedProportionFair}) into last term of right-hand-side of (\ref{DPPTheorem1}) with
\emph{slater-type conditions} \cite{neely2010stochastic},
and applying $\boldsymbol{\mu}^{*,\beta}(t) = \beta\mathbb{E}\{\boldsymbol{\mu}^{*,0}(\alpha^{\omega_j})\}$ to the right-hand-side of (\ref{DPPTheorem1}), we have
\begin{eqnarray}\label{SlaterTypeDPP}
  \nonumber &  
  \Delta(\boldsymbol{Q}(t)) - V\mathbb{E}\{\phi(\boldsymbol{\mu}(t))|\boldsymbol{Q}(t)\} \\
  \leq & 
  \mathcal {K} - V\phi_{\epsilon} -
  \beta \epsilon \sum_{u \in \mathcal {U}}\sum_{f \in \mathcal {F}}Q_{uf}(t)
\end{eqnarray}
thus we prove \emph{Lemma} \ref{l7}.
\IEEEQED

\section{}
\label{Theorem-2-proof}
First, we define the following optimization problem
\begin{equation}\label{DualOptimizationProblem}
  \text{max} \ \ g(\boldsymbol{\ell}),  \ \ \ \ \text{s.t.} \ \boldsymbol{\ell} \succeq 0,
\end{equation}
where $g(\boldsymbol{\ell})$ is called the dual function with the objective of original problem scaled by $V$. $\boldsymbol{\ell} = [q_1,\ldots,q_U,g_1,\ldots,g_U]$ is the Lagrange multiplier. $g(\boldsymbol{\ell})$ is defined as below,
\begin{eqnarray}
 \nonumber  g(\boldsymbol{\ell}) &=& \sum_{\omega_{j}}\pi_{\omega_{j}} \operatorname*{\text{inf} }\limits_{\boldsymbol{\mu}(\alpha_m^{\omega_{j}})} \ \Big\{ V\phi\big(\sum_{m}\varphi_m^{\omega_j}\boldsymbol{\mu}(\alpha_m^{\omega_{j}}) \big)\\
  \nonumber && + \sum_{u \in \mathcal {U}}q_u[\lambda_u - \sum_{m}\varphi_m^{\omega_{j}}\mu_u(\alpha_m^{\omega_{j}})], \label{DualFunction}
\end{eqnarray}
where we define the state space of $\omega(t)$ by $\overline{\Omega}= \{\omega_1,\omega_2,\cdots,\omega_J\}$, $\pi_{\omega_j}$ as the probability that $\omega(t) = \omega_j,j= 1,\cdots,J$, and the control action under the $\omega_j \in \overline{\Omega}$ as $\alpha_m^{\omega_j}$ with probability $\varphi_m^{\omega_j}$, where $\sum_{m} \varphi_m^{\omega_j} = 1$ and  $\varphi_m^{\omega_j} \geq 0$.

Let $\boldsymbol{\ell}^*$ denote the optimal solution of (\ref{DualOptimizationProblem}) and $\boldsymbol{\ell}^*$ is either $\Gamma(V)$ or zero. 
Then by \cite{Huanglongbo2011TAC}, we have the following lemma.
\begin{lemma}
\textit{
Suppose that
\begin{enumerate}
	\item The dual function $g(\boldsymbol{\ell})$ satisfies:
	\begin{equation}\label{DualFuncConstraint}
  g(\boldsymbol{\ell}^*) \geq g(\boldsymbol{\ell}) + L\parallel \boldsymbol{\ell}^*- \boldsymbol{\ell} \parallel, \ \ \ \ \forall \boldsymbol{\ell} \succeq 0,	
\end{equation}
for some constant $L > 0$ independent of V.
	\item The $\theta-\text{slack}$ condition (\ref{PredictiveSlaterCondition1}) is satisfied with $\theta > 0$.
\end{enumerate}
Then there exist constants $G$, $K$, $c$ such that for any $m \in \mathbb{R}_+$,
\begin{equation}\label{Theorem4Ineq}
  \mathcal {P}_r(G, Km) \leq ce^{-m}.
\end{equation}}
\textit{\noindent
We define $Q_u(t) \triangleq \sum_{f}\sum_{d = -1}^{D_u-1}\tilde{Q}_{uf}^d(t)$ for all $u \in \mathcal {U}$, and
\begin{equation}\label{TheoremEqu}
  \!\!\!\!\!\!\!\!
  \mathcal {P}_r(G, Km) \triangleq \limsup_{t \rightarrow \infty} \frac{1}{t} \sum_{\tau = 0}^{t-1}Pr \{\exists u, |Q_u(\tau)- q_u^*|\!\! >\!\! G + Km \}.
\end{equation}}
\end{lemma}
Based on Lemma 8, we further suppose that the dual function $g(\boldsymbol{\ell})$ satisfies
\begin{equation}\label{DualFuncConstraint}
  g(\boldsymbol{\ell}^*) \geq g(\boldsymbol{\ell}) + L\parallel \boldsymbol{\ell}^*- \boldsymbol{\ell} \parallel, \ \ \ \ \forall \boldsymbol{\ell} \succeq 0,
\end{equation}
for some positive constant $L_0$ independent of V, and the $\theta$-{slack} condition (\ref{PredictiveSlaterCondition1}) is satisfied with $\theta > 0$. 
Given such conditions, if FIFO queueing discipline is adopted and $D_u = O\big(\frac{1}{A_{\text{max}}}[q_u^* - G -K(\text{log}(V))^2 - \mu_{\text{max}}]^+ \big)$ for each user $u \in \mathcal {U}$, then by applying the proof techniques in \cite{Huang2014When}, it follows that 
PUARA can achieve an average queue backlog size reduction by at most $\sum_{u \in \mathcal {U}} D_u\big[\lambda_u - O(\frac{1}{V^{log(V)}}) \big]^+$.
\IEEEQED

\end{document}